\documentclass[12pt]{iopart}
\pdfoutput=1
\expandafter\let\csname equation*\endcsname\relax
\expandafter\let\csname endequation*\endcsname\relax 
\usepackage{mathtools} % which loads amsmath as well.
\usepackage[utf8x]{inputenc}
\usepackage{epsfig}
\usepackage{color}
\usepackage[toc,page]{appendix}
\usepackage{float}
\usepackage{graphicx}
\usepackage{anysize}
\usepackage{xfrac}
\usepackage{soul}
\soulregister\cite7
\soulregister\ref7
\soulregister\pageref7

\usepackage{soul}

\marginsize{3cm}{3cm}{2.5cm}{4.5cm}
\begin{document}

\title[Granular Motor in the Non-Brownian Limit]{Granular Motor in the Non-Brownian Limit}

\author{Loreto Oyarte Gálvez, Devaraj van der Meer }
\address{Physics of Fluids, Universiteit Twente, Post Office Box 217, 7500AE Enschede, The Netherlands}
\ead{l.a.oyartegalvez@utwente.nl}
\vspace{10pt}
\begin{indented}
\item[]February 2016
\end{indented} 

\begin{abstract} 
In this work we experimentally study a granular rotor which is similar to the famous Smoluchowski-Feynman device and which consists of a rotor with four vanes immersed in a granular gas. Each side of the vanes can be composed of two different materials, creating a rotational asymmetry and turning the rotor into a ratchet. When the granular temperature is high, the rotor is in movement all the time, and its angular velocity distribution is well described by the Brownian Limit discussed in previous works. 
When the granular temperature is lowered considerably we enter the so-called Single Kick Limit, where collisions occur rarely and the unavoidable external friction causes the rotor to be at rest for most of the time. We find that the existing models are not capable of adequately describing the experimentally observed distribution in this limit. We trace back this discrepancy to the non-constancy of the deceleration due to external friction and show that incorporating this effect into the existing models leads to full agreement with our experiments. 
Subsequently, we extend this model to describe the angular velocity distribution of the rotor for any temperature of the gas, and obtain a very good agreement between the model and experimental data. 
\end{abstract}
 
%%%%%%%%%%%%%%%%%%%%%%%%%%%%%%%%%%%%%%%%%%%%%%%%%%%%%%%%%%%%%%%%%%%%%%%%%%%%%%%%%%%%%%%%%%%%%%%%%%%
%%%%%%%%%%%%%%%%%%%%%%%%%%%%%%%%%%%%%%%%%%%%%%%%%%%%%%%%%%%%%%%%%%%%%%%%%%%%%%%%%%%%%%%%%%%%%%%%%%%

\section{Introduction}

The attempts to challenge the second law of thermodynamics have been many throughout history. In 1912, Marian Smoluchowski devised a prototype, consisting of a rotor combined with a ratchet and pawl, designed to convert the Brownian motion of the rotor into work (Fig.~\ref{fig:ExperimentalSetup}-a)\cite{smoluchowski_experimentell_1927}. Fifty years later Feynman showed unambiguously why at thermal equilibrium this device cannot actually do this \cite{feynman_feynman_1977}, firmly establishing the validity of the second law. However, far from equilibrium, the behaviour of a rotor which rectifies motion of randomly moving molecules in their surroundings is still an active matter of study. These so-called molecular motors are believed to be responsible for tensing and relaxing the muscles of the body, for numerous cellular and intracellular transport process, photovoltaic and photorefractive effects, among many other processes \cite{reimann_brownian_2002,talbot_effect_2012,sarracino_ratchet_2013,hanggi_brownian_2005}.

A granular motor can be obtained by immersing a rotor very similar to the one of the Smoluchowski-Feynman device into a granular gas. This rotor can turn freely due to (dissipative) collisions with the gas particles; for a symmetric rotor this motion will be symmetric but as soon as the symmetry is broken --owing to the fact that the system is far from thermal equilibrium-- the rotor will turn in a preferred direction and therefore starts operating as a motor, much like the device envisioned by Smoluchowski would have done  \cite{eshuis_experimental_2010,balzan_brownian_2011}. 

In an experimental setup the rotor will naturally experience external friction in the bearings that connects its axis to rest of the experimental setup. As a result 
we can distinguish two limiting behaviours depending on how frequent collisions with the rotor occur: We will denote these as the Brownian Limit and the Single Kick Limit \cite{talbot_kinetics_2011,gnoli_granular_2013} respectively. In the Brownian Limit the collisions occur very frequently such that the rotor remains in motion all the time and dissipation due to external friction in between two kicks is negligible. In contrast, in the Single Kick Limit the collisions occur so rarely that due to the external friction the rotor is typically able to relax its velocity to zero after each kick and remains in rest until the next kick occurs. This second limiting regime clearly can not exist without external friction.     
  
Several studies strived after understanding and modeling the granular motor in these limits, which has lead to an adequate description in the Brownian Limit \cite{cleuren_dynamical_2008,eshuis_experimental_2010,talbot_kinetic_2011,talbot_kinetics_2011,joubaud_fluctuation_2012}. However, the few theoretical studies that exist in the Single Kick Limit compare well with particle simulation but do not have a good agreement with experimental results \cite{talbot_kinetics_2011,gnoli_granular_2013,kanazawa_asymptotic_2014}. Moreover, analysing the behaviour of the rotor in between these two limits appears to bea very hard problem to address in general.     
       
In this work, we focus on experimentally studying the behaviour of the rotor both in the Single Kick Limit and beyond, going towards the Brownian Limit. Subsequently, we will construct a model that is valid for both situations by taking into account two important considerations: First, we will consider the fact that the external friction plays an important role. And secondly, we assume that the statistics of the kicks that the rotor experiences and the subsequent deterministic velocity decrease due to external friction are mutually independent. Then, we analyse the friction effects on the rotor relaxation after a kick to obtain a model for its angular velocity distribution. Finally, we compare our model with the experimental results and obtain a very good agreement between them.

This article is organised as follows. In Section \ref{s:expSetup}, the experimental setup is detailed and the different limiting behaviours are defined in greater depth. In Sections \ref{s:SKL} and \ref{s:B-SKL}, the model to describe the angular velocity distribution is developed, for the Single Kick Limit and beyond the Single Kick Limit respectively. In addition, Sections \ref{s:SKL} and \ref{s:B-SKL} include the comparison between the model and the experimental results. Finally, in section \ref{s:Conclusions} a summary of this study is presented.

%%%%%%%%%%%%%%%%%%%%%%%%%%%%%%%%%%%%%%%%%%%%%%%%%%%%%%%%%%%%%%%%%%%%%%%%%%%%%%%%%%%%%%%%%%%%%%%%%%%
%%%%%%%%%%%%%%%%%%%%%%%%%%%%%%%%%%%%%%%%%%%%%%%%%%%%%%%%%%%%%%%%%%%%%%%%%%%%%%%%%%%%%%%%%%%%%%%%%%%

\section{Experimental Setup and Limiting Behaviours}\label{s:expSetup}

\begin{figure}[h!]
  \begin{center}
   \includegraphics[scale=0.25]{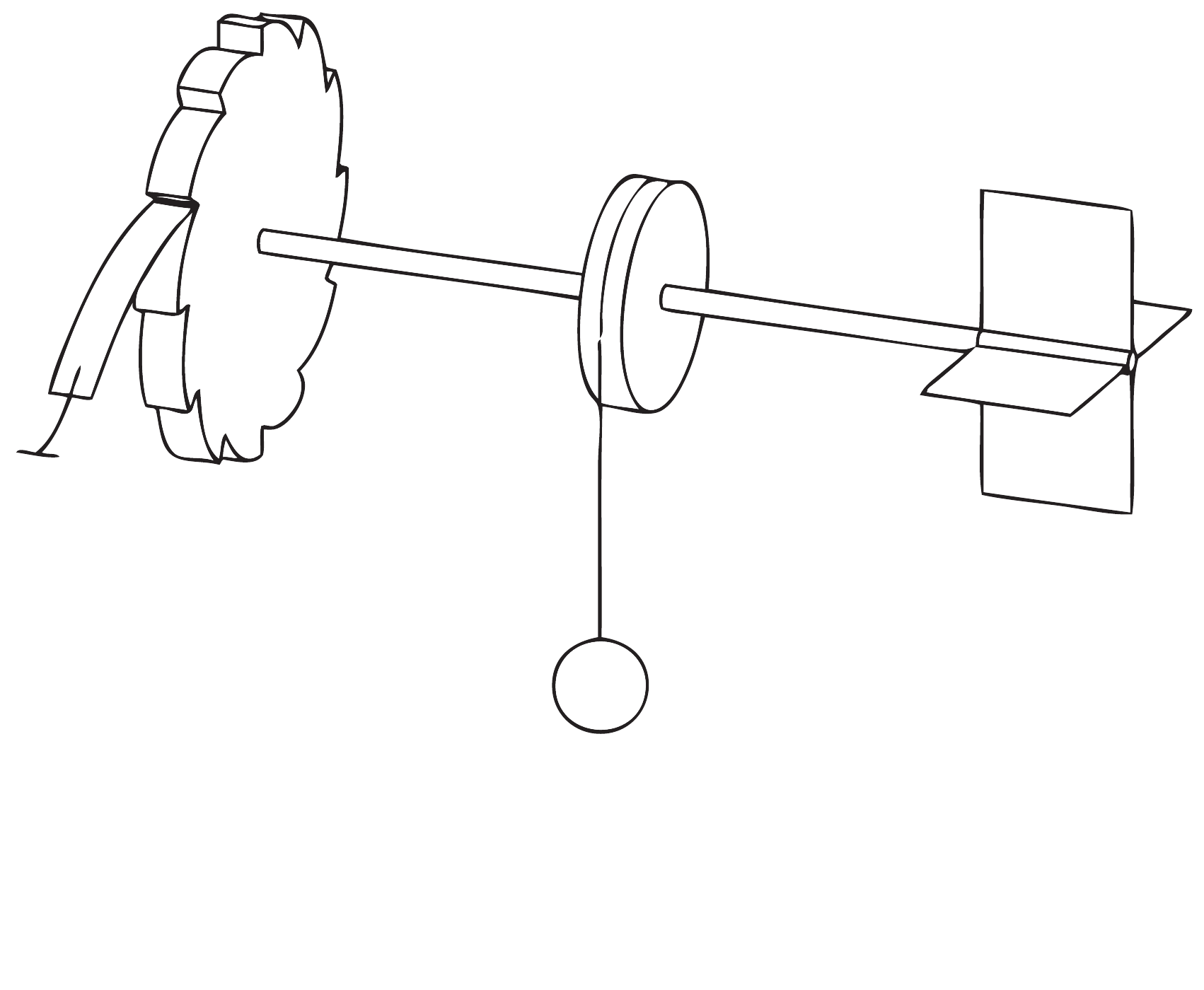}\qquad
   \includegraphics[scale=0.6]{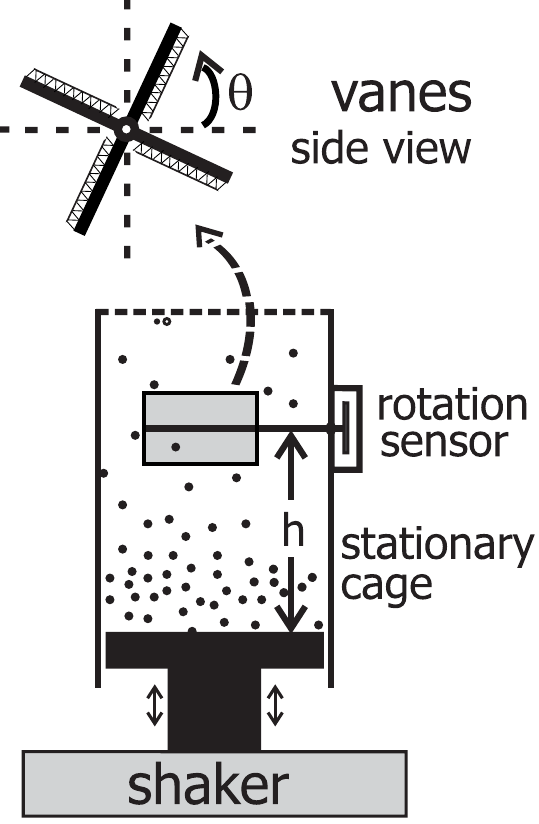}\qquad
  \includegraphics[scale=0.4]{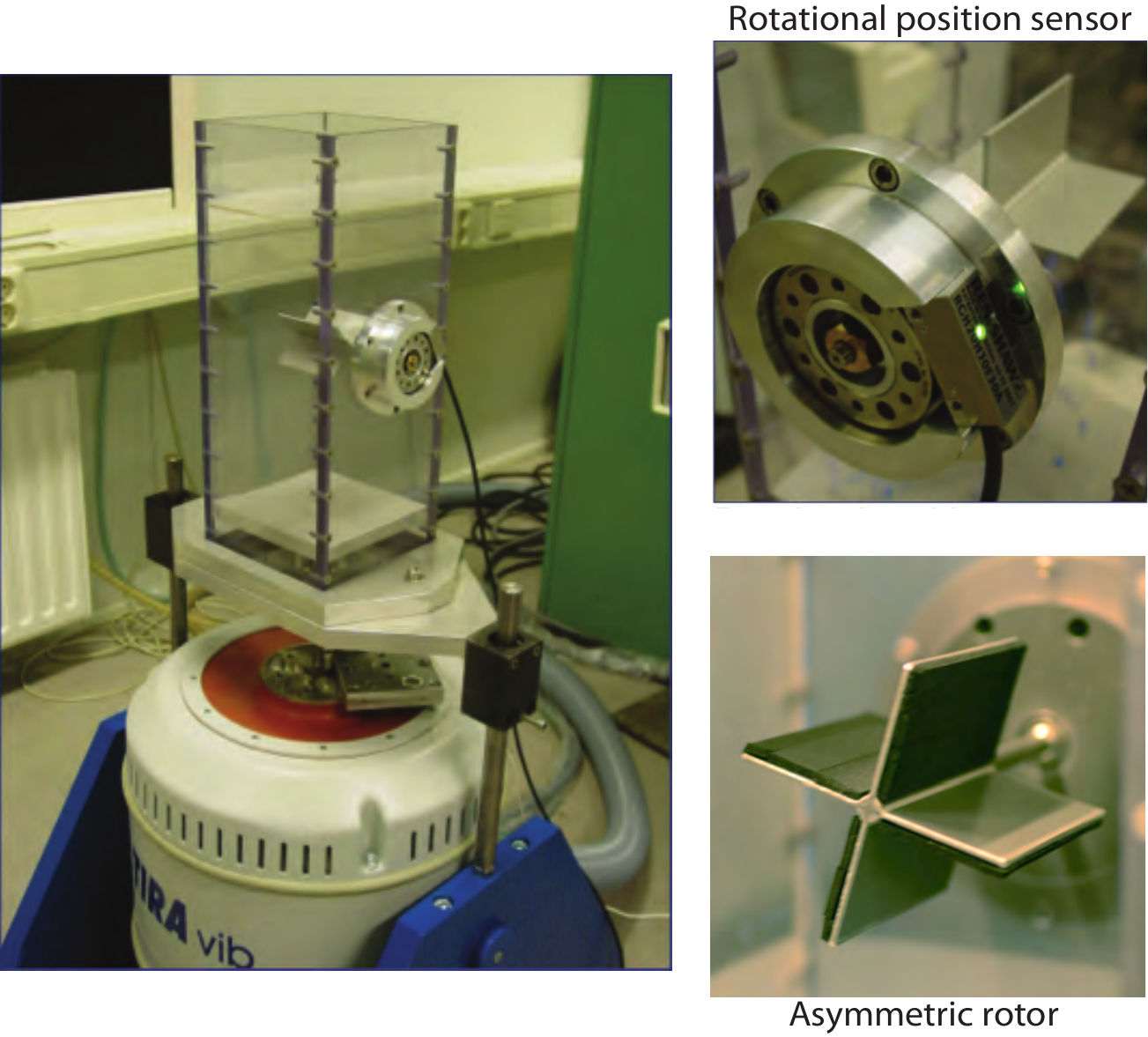}\\
  \hspace{-0.5cm}(a) \qquad\qquad\qquad\qquad(b) \qquad\qquad\qquad\qquad\qquad \quad (c)
  \end{center}
  \caption{(a) Schematic of the Smoluchowski-Feynman device with ratchet and pawl on the left and rotor on the right. (b) Schematic of the experimental setup. A rotational position sensor is fixed in the wall where the rotor is connected, measuring its angular position in time $\theta(t)$. Side view of the vanes showing different material on each side implying a different coefficient restitution $\alpha_-$ (for kicks resulting in anti-clockwise motion) and $\alpha_+$ (clockwise). (c) Experimental setup in the lab.}
  \label{fig:ExperimentalSetup}
 \end{figure}

In order to study a granular rotor, we built a setup consisting of an acrylic container with the objective to confine a granular gas, i.e., preventing the particles from leaving the system, as shown in Fig.\ref{fig:ExperimentalSetup}. 

The granular gas is formed by $N_p = 20$ steel spheres of diameter $d = 10$ mm and density $\rho=7.8$ g/cm$^3$. 
They are brought into a gas-like state by a vibrating bottom, which is mounted on a shaker with tuneable frequency $f$ and amplitude $a$. The distance between the bottom and the axis is fixed to $h = 51$ mm. Thus, the container is a stationary perspex cage in which the vibrating bottom wall is moving like a piston. It is important to note that the air pressure inside the container is constant and of no influence on the motion of grains and rotor. \cite{eshuis_experimental_2010} 

The rotor is composed of four vanes ($30\times60$ mm$^2$ each, made from one piece of stainless steel) that are precisely balanced around an axis, which in turn is connected to the container wall by a low-friction ball bearing. The angle $\theta(t)$ is measured by an optical angle encoder and the acquisition frequency is set to $2,000$ Hz, thus providing the angular velocity $\Omega(t)$ of the rotor at any time.

To obtain a granular Brownian motor, the symmetry of the rotor is broken by mounting on the right side of each vane a neoprene sealing strip, with 2 mm thickness. In this way, the coefficient of normal restitution ($\alpha$) is diminished on one side with respect to the other, and by that the energy dissipated after a kick will be different on each side, inducing a ratchet effect \cite{costantini_models_2009} working similarly as the ratchet and the pawl in the Smoluchowski-Feynman device of Fig.~\ref{fig:ExperimentalSetup}-a.

Changing the properties of the granular gas particles, the frequency $f$ and amplitude $a$ of the shaker or the number of particles $N_p$, the collisions between particle-vane become more or less frequent. Here we choose to vary the frequency and depending on this frequency we can distinguish the two limiting behaviours introduced before, namely the Brownian and the Single Kick Limits.  

\begin{figure}[h!]
  \begin{center}
  \includegraphics[scale=0.25]{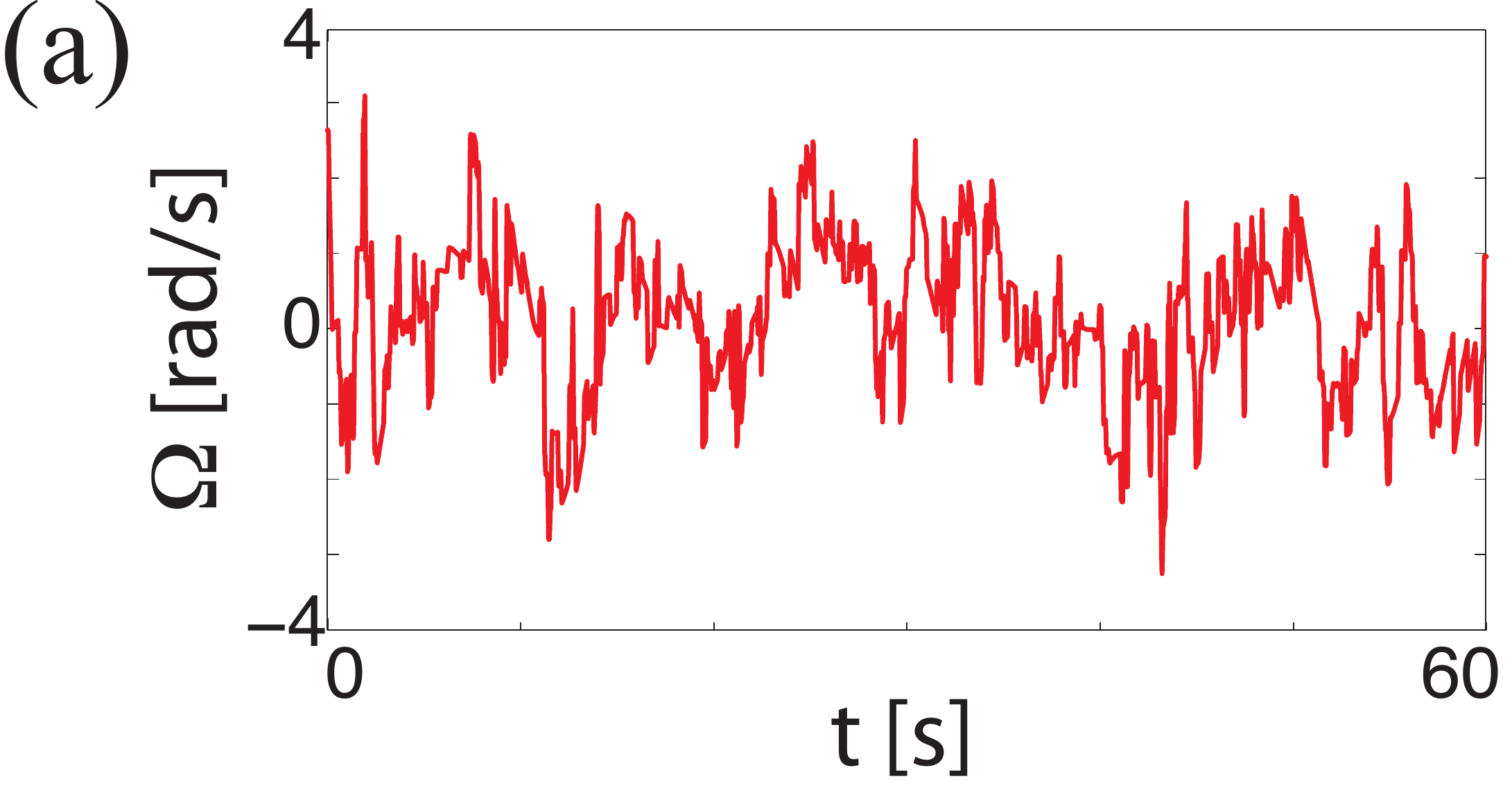}\qquad\qquad
  \includegraphics[scale=0.25]{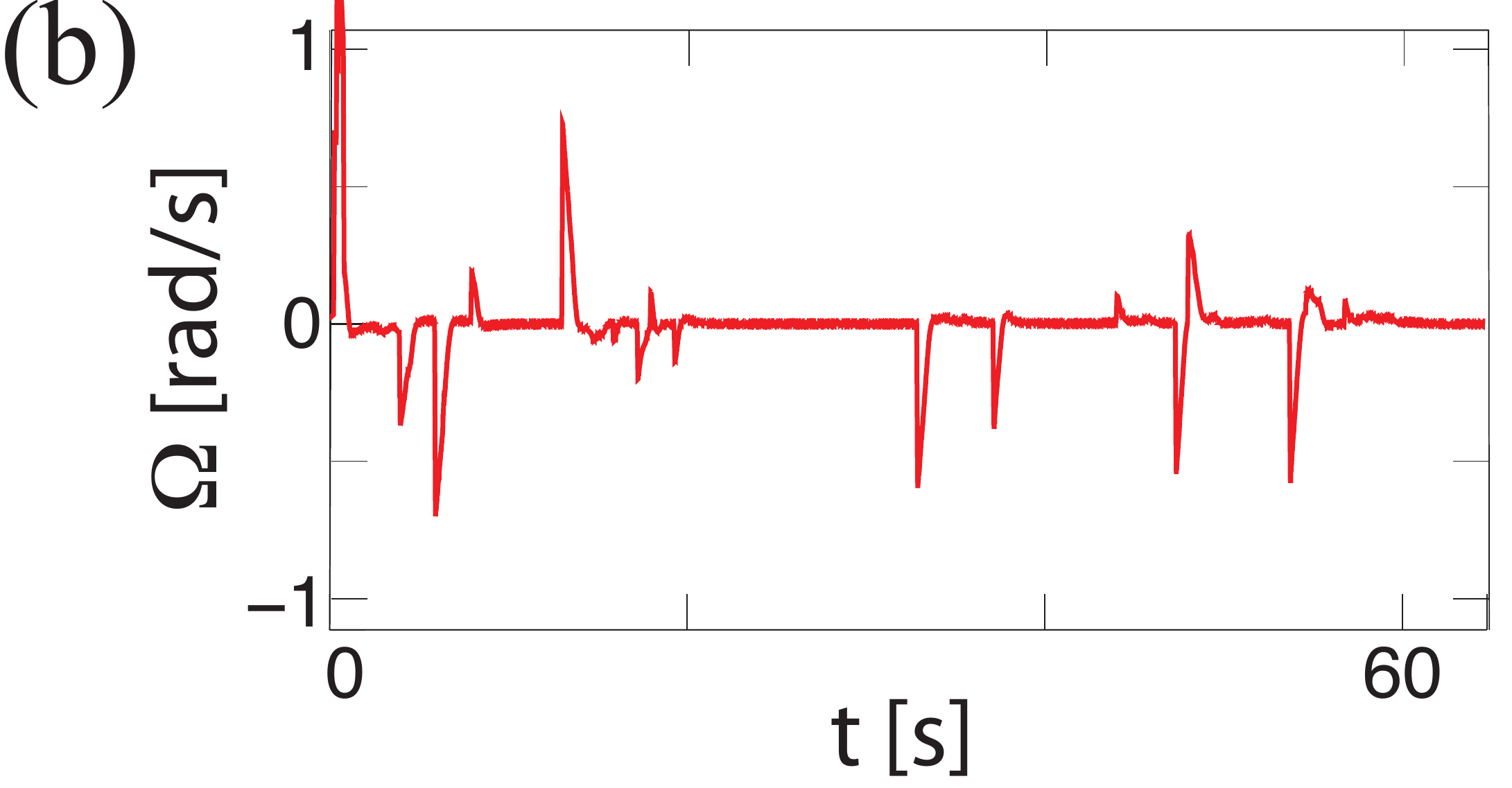}
  \end{center}
  \caption{Time evolution of the angular velocity of the rotor for (a) the Brownian Limit (for $f=40$ Hz and $a=1.4$ mm) and (b) the Single Kick Limit (for $f=20$ Hz and $a=1.4$ mm).}
  \label{fig:LimitingBehaviors}
\end{figure}

In Fig.~\ref{fig:LimitingBehaviors}-a we plot the typical time evolution of the angular velocity of the rotor in the Brownian Limit: Here one observes that the particles-vanes collisions are very frequent; before the rotor can start to relax noticeably immediately another kick occurs due to which the rotor is in motion all the time. 
In this limit, the behaviour of the rotor --most specifically the angular velocity distribution-- is well described from both a theoretical and a numerical perspective, and with good agreement with experimental results \cite{cleuren_dynamical_2008,eshuis_experimental_2010,talbot_kinetic_2011,talbot_kinetics_2011,joubaud_fluctuation_2012}. 

In the Single Kick limit, the energy injected is low, the gas is very diluted (composed of only few particles) and hence the particle-vane collisions are not frequent; only occasionally a particle-vane collision sets the vanes into motion. Whenever a kick occurs the rotor has time to fully relax under the influence of the external friction and will stay in rest until the next collision; in this limit the rotor is in rest for most of the time. This behaviour shows up as many isolated peaks in the time evolution of the angular velocity, as plotted in Fig.~\ref{fig:LimitingBehaviors}-b.  
Note that this limit would not be possible without external friction, because it is this friction that is responsible for the relaxation of the rotor after a kick. There exist a few studies that describe the angular velocity distribution of the rotor in this limit, both theoretically and numerically \cite{talbot_effect_2012,talbot_kinetic_2011,kanazawa_asymptotic_2014}, but they are not in  agreement with experimental results \cite{gnoli_granular_2013}.

From the above description of the limiting behaviours, it becomes clear that there are two relevant time scales present in the system: The relaxation time $\tau_s$, corresponding to the average stopping time of the rotor, due to external friction, and the collision time $\tau_c$, corresponding to the average time between particle-vane collisions. In the Brownian Limit the relaxation time is much larger than the collision time $(\tau_s\gg\tau_c)$, whereas in the Single Kick Limit the relaxation time is much smaller than the collision time $(\tau_s\ll\tau_c)$. 

In the next sections we will develop a model to describe the angular velocity distribution (AVD) of the rotor, starting with the Single Kick Limit (Section \ref{s:SKL}) and subsequently moving beyond this limit, towards the Brownian Limit (Section \ref{s:B-SKL}). We show what role the external friction plays in the relaxation of the rotor and how it has to be incorporated into the model to obtain a good agreement with the experimental data.    

%%%%%%%%%%%%%%%%%%%%%%%%%%%%%%%%%%%%%%%%%%%%%%%%%%%%%%%%%%%%%%%%%%%%%%%%%%%%%%%%%%%%%%%%%%%%%%%%%%%%%%%%%%%%%%%%%%%%%%%%%%%%%%%%%%%%%%%%%%%%%%%%%%%%%%%%%%%%%%%%%%%%%%%%%%%%%%%%%%%%%%%%%%%%%%%%%%%%

\section{Single Kick Limit}\label{s:SKL}

\begin{figure}[h!]
  \begin{center}
   \includegraphics[scale=0.30]{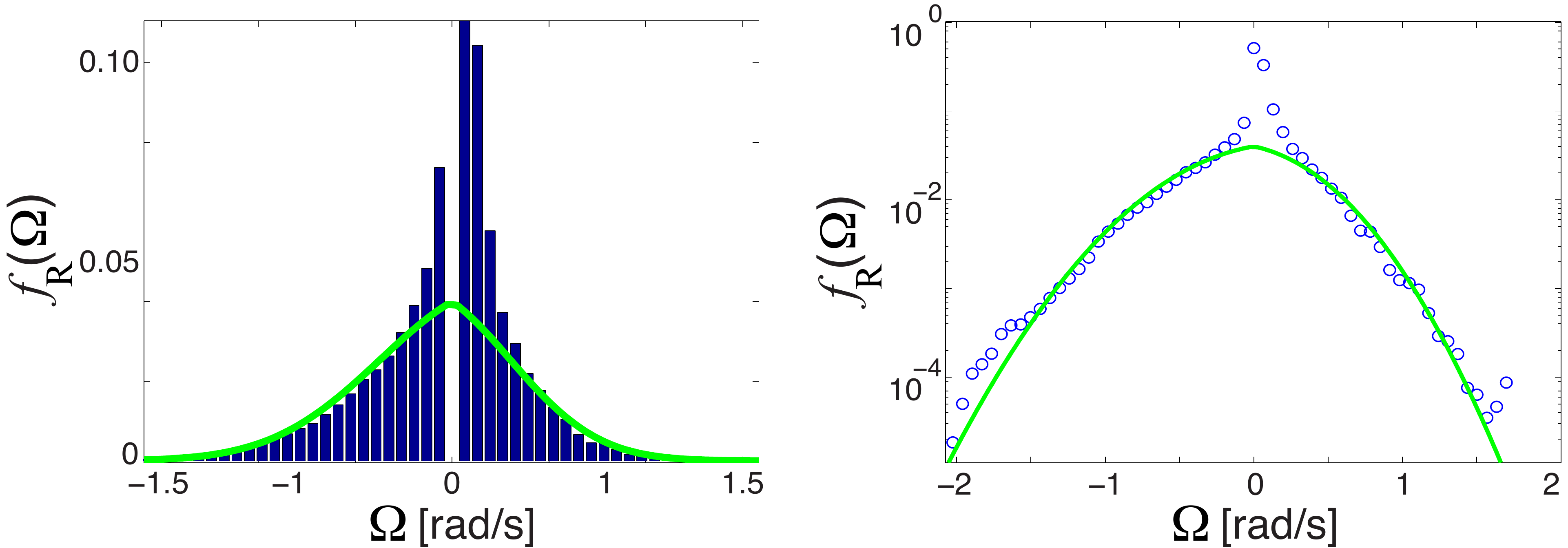}  
  \end{center}
  \caption{Regular part of the experimentally determined angular velocity distribution (AVD) in the Single Kick Limit for a typical non-symmetric case ($f$=20Hz, $a$=2.8mm), the vertical axes on the left and right plots are linear and logarithmic respectively. The green line corresponds to the model from Talbot \textit{et al.} fitted to the experimental data. Clearly, whereas the fit is reasonable for larger velocities it fails to describe the experiments for velocities closer to zero.}
  \label{fig:AVDKickRegimeTV}
  \end{figure}

As stated before, there are few studies that have addressed the Single Kick Limit, and the agreement between the theoretical/numerical work on the one side and the experimental work on the other is not satisfactory. Talbot \textit{et al.} \cite{talbot_kinetics_2011} were the first to develop a model for the AVD of the rotor in the Single Kick Limit. The distribution shows a non-Gaussian shape, contrary to the Brownian Limit \cite{eshuis_experimental_2010,joubaud_fluctuation_2012}, and fits very well to numerical simulations. The same is true for  the experimental data for large velocities, but the model does not work for velocities close to zero, as is seen in the Fig.~\ref{fig:AVDKickRegimeTV} where we compare our experimental data to the model of \cite{talbot_kinetics_2011}.  The same holds for earlier experiments performed in this regime \cite{gnoli_granular_2013} which could also not be fitted with the model from \cite{talbot_kinetics_2011}.

\begin{figure}[h!]
 (a)\hspace{7cm}(b)\vspace{-4 mm}
\begin{center}
\includegraphics[scale=0.25]{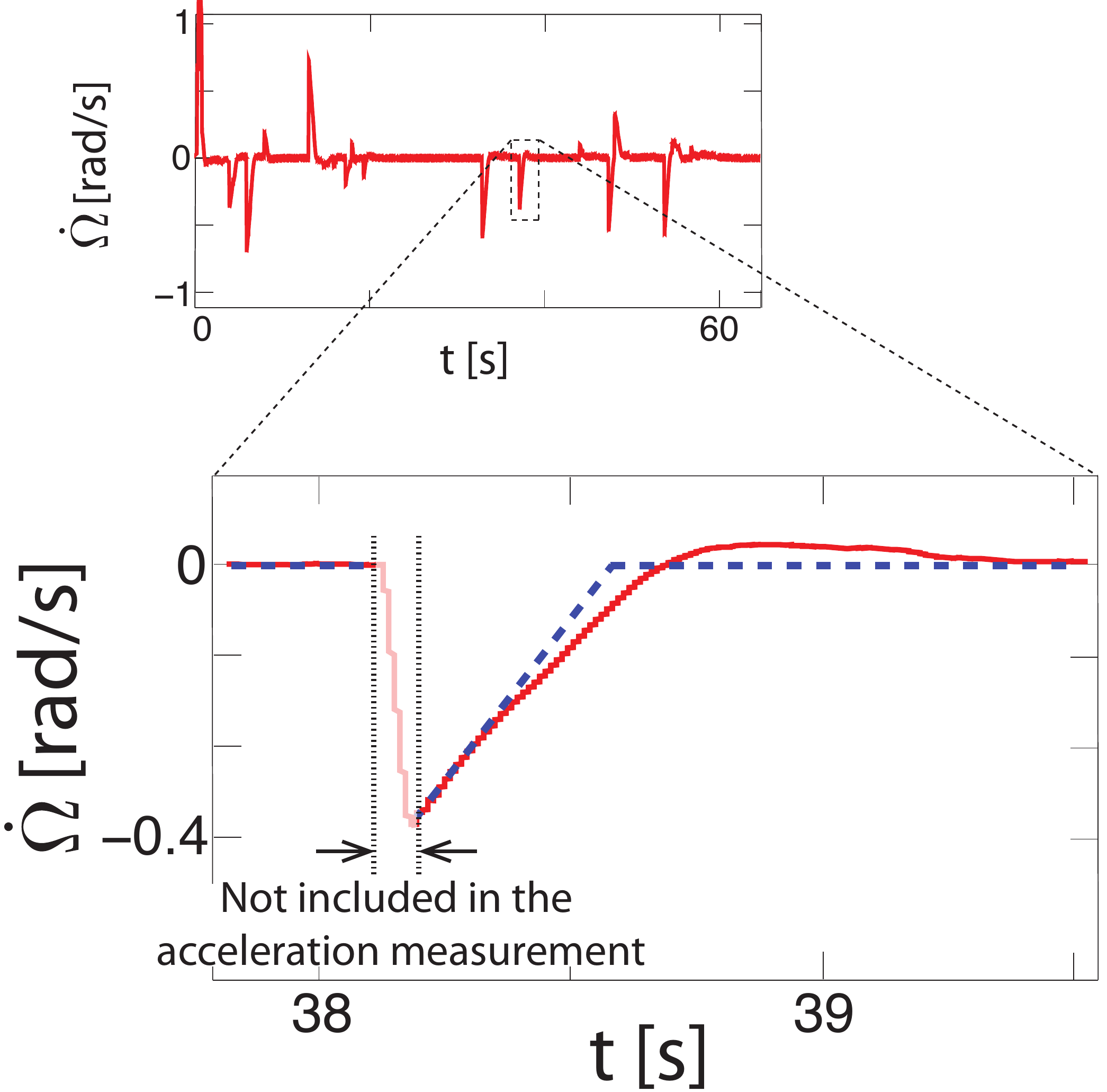}\qquad\qquad 
\includegraphics[scale=0.4]{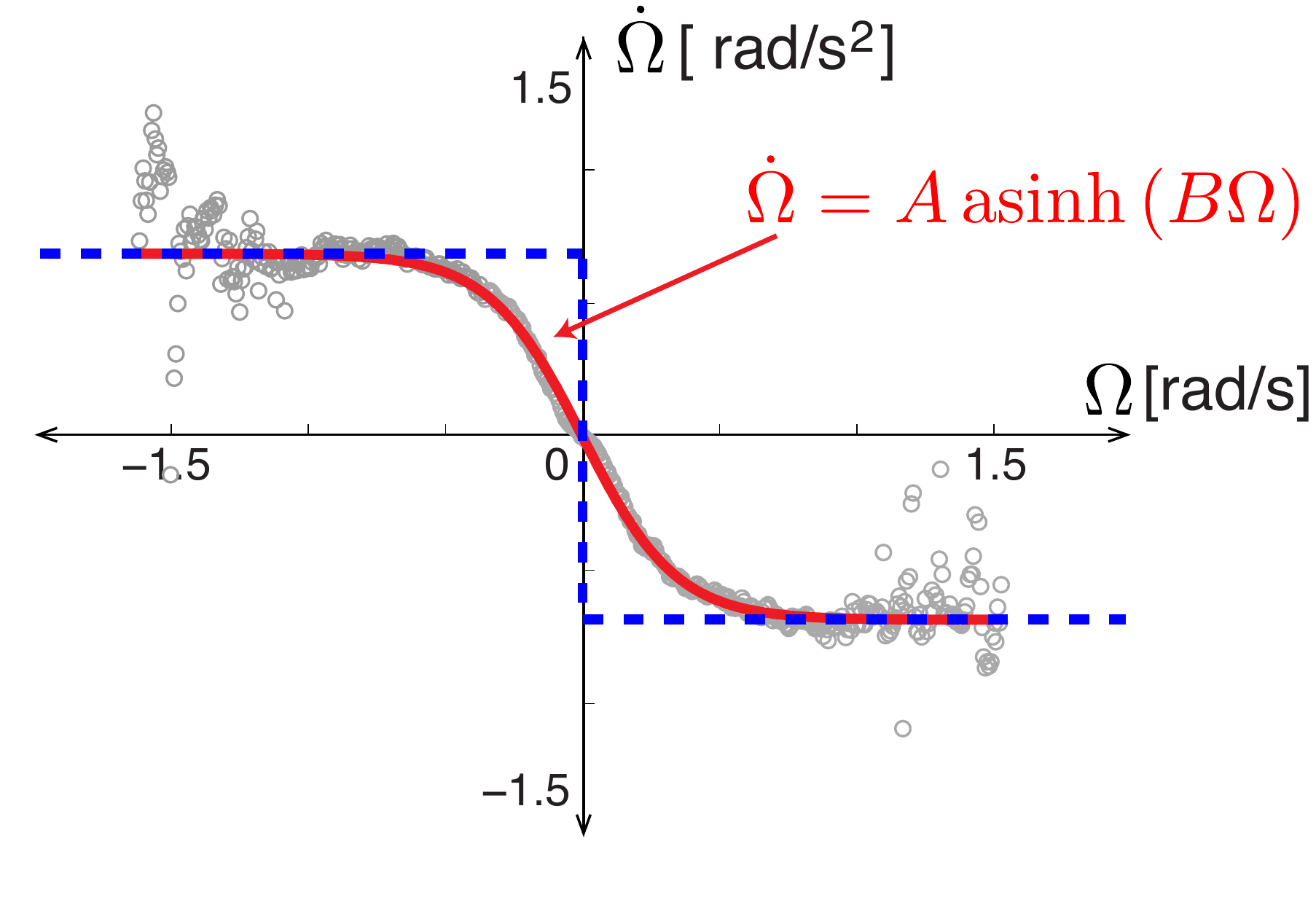} \vspace{3 mm}
\end{center}
(c)\vspace{-5 mm}
\begin{center}
\includegraphics[scale=0.43]{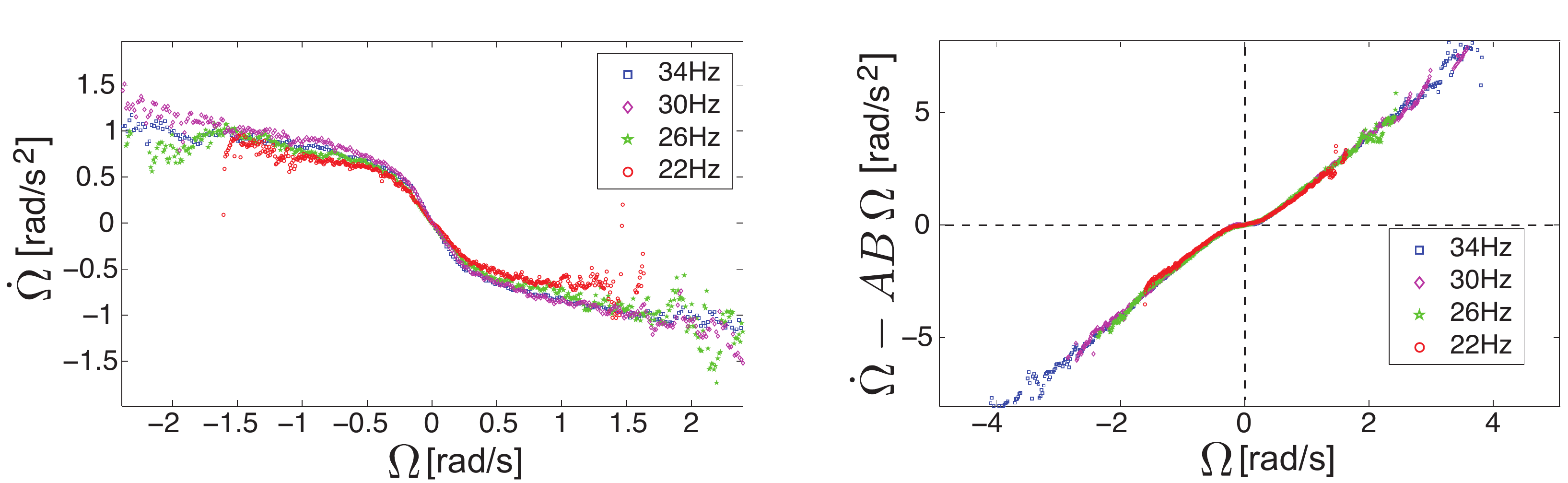}
\end{center}
\caption{(a) Zoomed in view of the angular velocity signal, after a kick. The red line are the experimental data and the blue dashed line represents the theoretical angular velocity when the external friction is constant. (b) Angular acceleration $\dot{\Omega}$ versus angular velocity $\Omega$ during the relaxation of the rotor after a kick has occurred. The grey circles refer to experimental data, while the red line is a phenomenological equation that fits the data, and the blue dashed-line correspond to an angular acceleration with constant external friction. (c) Angular acceleration versus angular velocity for different frequencies $f=22$, 26, 30 and 34 [Hz] {(both in and beyond the Single Kick Limit)} are plotted on the left. On the right, the first order term from the fit is subtracted from the angular acceleration, showing the data converge in a curve independent of the frequency.}
\label{fig:AvsV}
\end{figure}

One of the key ingredients for the model of Talbot \textit{et al.} and one of the possible differences between the particle simulations and the experiments is that the external friction affecting the rotor is assumed to be constant. Therefore, we decided to test this assumption in our experiments by analysing the angular acceleration $\dot{\Omega}$ versus the angular velocity $\Omega$. In order to do this after every kick we determine $\dot{\Omega}$ from the $\Omega(t)$-curve until the rotor reaches rest, i.e, the acceleration data does not include the exact moment when the kick occurs, as shown by the faded line in Fig.~\ref{fig:AvsV}-a. This plot of $\dot{\Omega}$ versus $\Omega$ is presented in Fig.~\ref{fig:AvsV}-b, where the grey circles correspond to experimental data, and the blue dashed-line is the acceleration when the external friction is constant. Clearly, we observe a non-constant external friction in our experiments.  

We note that, incidentally, the experimental data is well fitted by a inverse hyperbolic sine, which is convenient to use in the following derivation of the model. Figure~\ref{fig:AvsV}-c (left), shows the the angular acceleration $\dot{\Omega}$ versus the angular velocity $\Omega$ for different frequencies, showing small differences between them when the frequency is increased. However, when we subtract the linear term from each fit (which is performed for each frequency separately) to the experimental data and plot the result versus angular velocity (Fig.~\ref{fig:AvsV}-c, right), we find that the data converge onto a single curve. Therefore, differences between the acceleration curves are likely to correspond to the stochastic motion of the particles in the gas.\footnote{\\The friction measured is not only the ball bearing friction, but includes the friction the rotor experiences due to its motion through the granular gas as well.} Because the differences are small we will neglect them in the following.

\begin{figure}[h!]
  \begin{center}
  \includegraphics[scale=0.5]{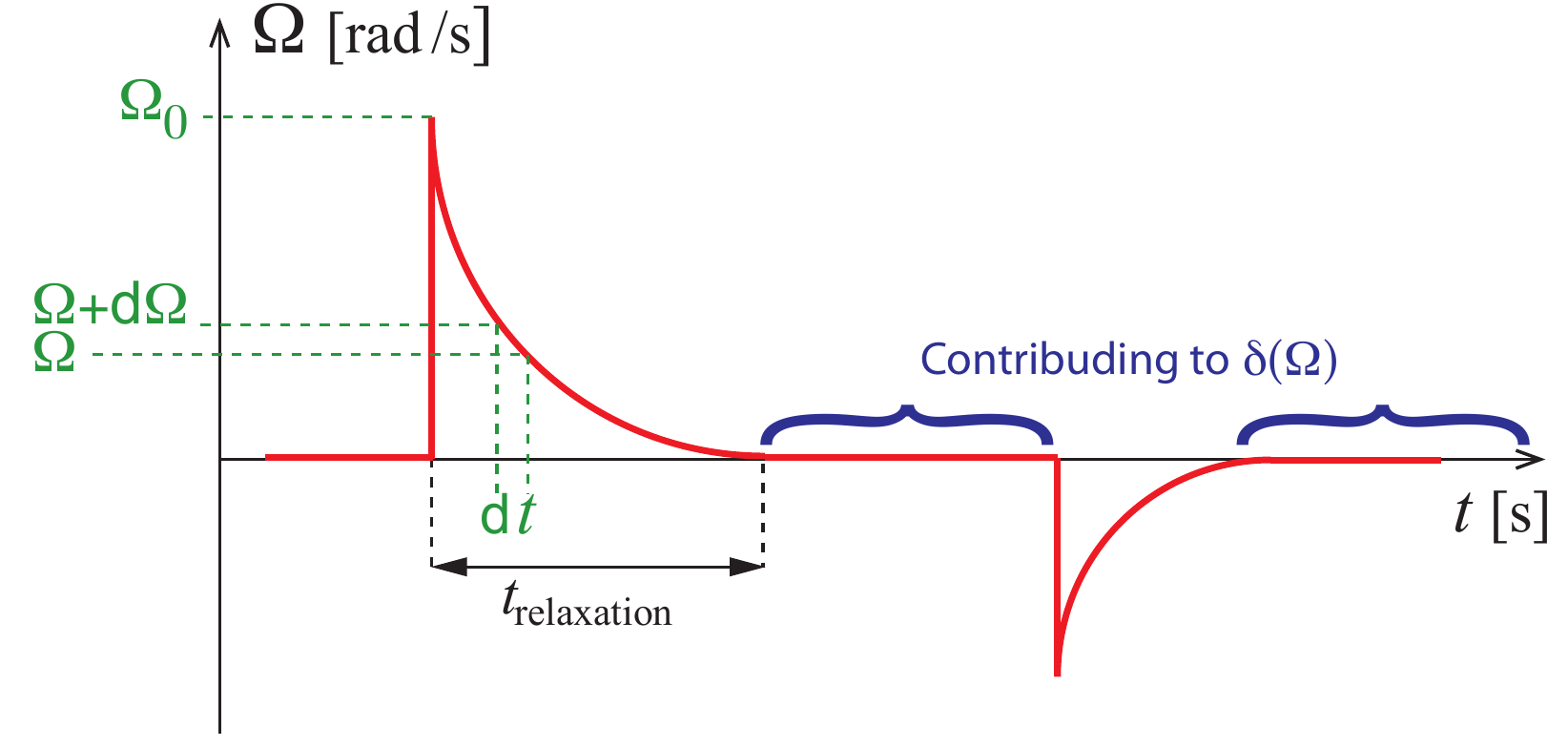}
  \end{center}
  \caption{A schematic representation of the time evolution of the angular velocity in the Single Kick Limit. Time intervals during which the rotor is at rest contribute to the singularity in the angular velocity distribution. For a kick with initial angular velocity $\Omega_0$, the time that the rotor takes to come to the rest is $t_\textrm{relaxation}$ and depends only of the initial velocity $\Omega_0$.}
  \label{fig:schematicKickRegime}
 \end{figure}

We now turn to the derivation of the model, where we closely follow that of Talbot \textit{et al.} in \cite{talbot_kinetics_2011}. From the plots in Fig.~\ref{fig:LimitingBehaviors}-b and Fig.~\ref{fig:schematicKickRegime} we observe that the Angular Velocity Distribution $f(\Omega)$ contains two parts: (i) a singularity $\delta\left(\Omega\right)$ corresponding to the time during which the rotor is at rest and (ii) a (normalised) regular part $f_{R}\left(\Omega\right)$ 
corresponding to the relaxation of the rotor after a kick, \textit{i.e.},

\begin{equation}
f\left(\Omega\right)=\gamma\delta\left(\Omega\right)+\left(1-\gamma\right)f_{R}\left(\Omega\right)\label{eq:AVD}
\end{equation}

$\fl\hspace{3.8cc}$where $\gamma$ is a normalisation constant {which may be determined from the conservation of probability current \cite{talbot_kinetics_2011} but in our case will be obtained as a fitting parameter from the experimental data}. 
 
Given this expression, we want to model the regular part of the AVD by analysing a single kick as shown in Fig.~\ref{fig:schematicKickRegime}. If we consider a particle-vane collision which gives the rotor an initial angular velocity $\Omega_0$, the time that the rotor needs to come to rest will be $t_\textrm{relaxation}$, which only depends on the initial velocity.  Therefore, we must consider two ingredients to define the regular part of the AVD: The distribution of the particle-vane kicks $G(\Omega_0)$ and the probability to find the rotor at a certain velocity between $\Omega$ and $\Omega+\delta\Omega$, given that the initial velocity after the kick is $\Omega_0$.

First, we need to provide the kick distribution depending on the particle velocity. To this end, we assume that the velocities of the particles in a granular gas follow a Maxwell distribution, $\Phi\left(v\right)=\sqrt{\frac{m}{2\pi T}}\textrm{e}^{-\frac{m}{2T}v^{2}}$ where $v$ is the normal velocity of a particle relative to the vane just before a kick, $m$, the mass of the particle and $T$, the granular temperature (defined as the average kinetic energy of the particles, after a possible mean flow has been subtracted). The probability $G(v)\,\textrm{d}v$ that a granular gas particle hits a vane with a normal velocity $v${, assuming a uniform distribution along the axis of the rotor, is computed perpendicular to the axis and} can be expressed as:

\begin{equation}
G\left(v\right)\,\textrm{d}v=\int_{-\frac{L}{2}}^{\frac{L}{2}}\textrm{d}x\,\rho\left|v\right|\Phi\left(v\right)\,\textrm{d}v
\label{eq:normveldist}
\end{equation}

\begin{figure}[h!]
  \begin{center}
  \includegraphics[scale=0.25]{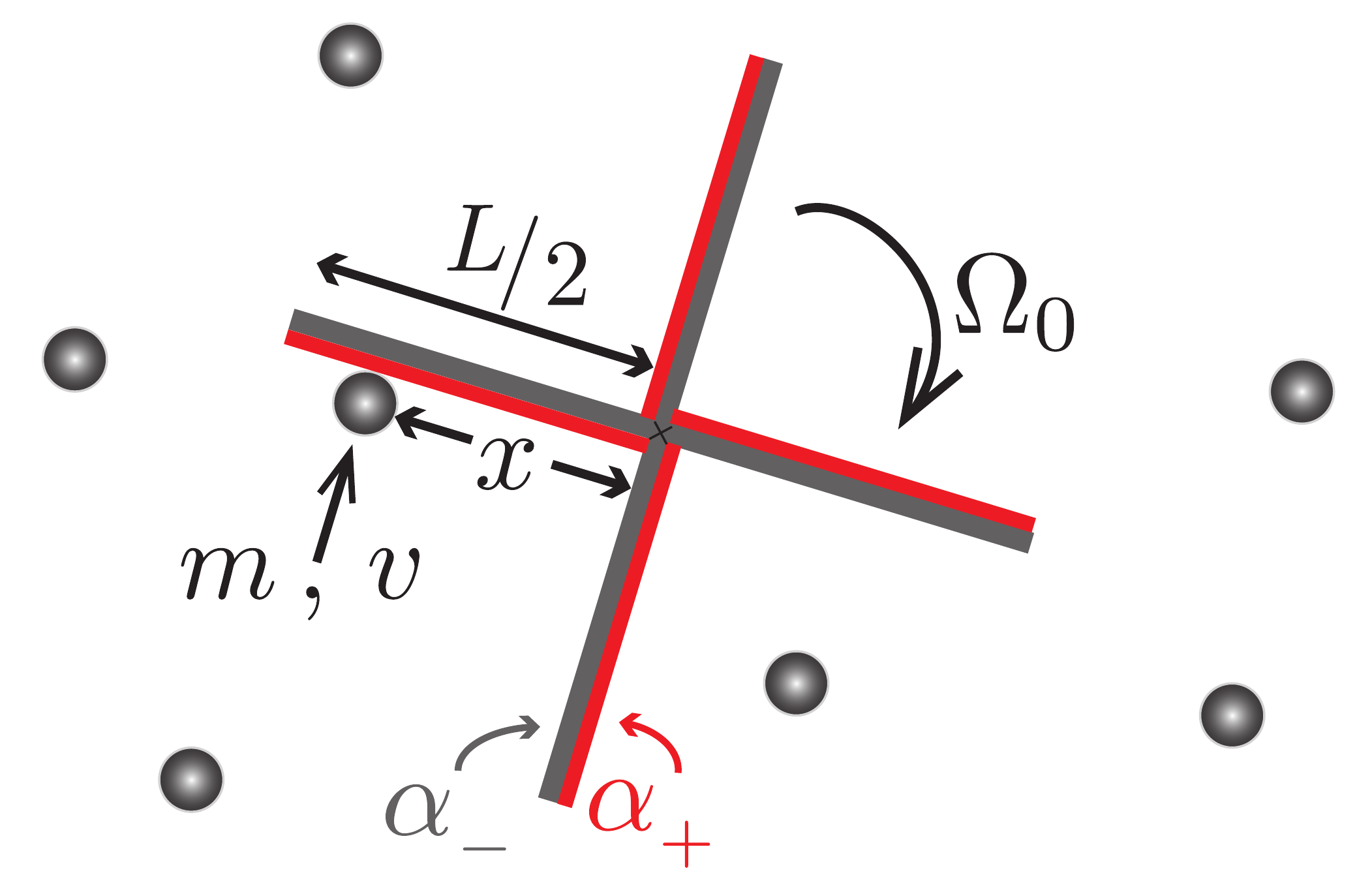}
  \end{center}
  \caption{Schematic of the particle-vane interaction, where the red and grey side of each vane may be composed of different materials, with corresponding coefficients of restitution $\alpha_+$ (clockwise, positive direction) and $\alpha_-$ (anticlockwise, negative direction) respectively.}
  \label{fig:momentum}
 \end{figure}

$\fl\hspace{3.8cc}$where $\rho$ is the density of the granular gas, $L$ the length of the vane {(as shown in figure \ref{fig:momentum})} and $v$ the normal velocity of a particle. 

By conservation of angular momentum, we can relate the normal particle velocity before a kick with the angular velocity immediately after a kick ($\Omega_0$), when the particle hits the vane at a position that corresponds to a distance $x$ from the axis, as shown in Fig.~\ref{fig:momentum}, 

\begin{eqnarray}
mxv & = & -mx\alpha_{\pm}v+mx^{2}\Omega_{0}+I\Omega_{0}\\
v & = & \frac{I+mx^{2}}{(1+\alpha_{\pm})mx}\Omega_{0}
\label{eq:momentum}
\end{eqnarray} 

$\fl\hspace{3.8cc}$where $I$ corresponds to the moment of inertia of the rotor, and $\alpha_-$ and $\alpha_+$ are the coefficients of restitution {on each side of the vane (see figure \ref{fig:momentum}).}
 
Therefore, using equations~\ref{eq:normveldist}-\ref{eq:momentum} and integrating over the distance $x$ to the axis, we obtain the dependence of the kick distribution on the angular velocity where we have introduced the variable changes $z=\frac{2x}{L}$ and $\xi=\frac{mL^{2}}{4I}$, leading to the following expression for the kick distribution $G_{\pm}(\Omega_{0})$ 

\begin{equation}
G_{\pm}\left(\Omega_{0}\right)\,\textrm{d}\Omega_0=\int_{-1}^{1}\textrm{d}z\,\frac{\left(1+\xi z^{2}\right)^{2}I}{T\left(1+\alpha_{\pm}\right)^{2}\xi z^{2}}|\Omega_{0}|\,\textrm{e}^{-\frac{\left(1+\xi z^{2}\right)^{2}I}{2T\left(1+\alpha_{\pm}\right)^{2}\xi z^{2}}\Omega_{0}^{2}}\,\textrm{d}\Omega_0
\label{eq:KickDistribution}
\end{equation}
 
The second ingredient necessary to obtain the regular part of the AVD is the probability $h\left(\Omega|\Omega_{0}\right)$ to find the rotor with an angular velocity between $\Omega$ and $\Omega+\mathrm{d}\Omega$ after a single kick has given it an initial velocity $\Omega_{0}$, as shown in Fig.~\ref{fig:schematicKickRegime}. {For clockwise rotation, this probability is the time dt between these two velocities over the total time the rotor needs to relax, $t_{relaxation}$, multiplied by the probability of finding the rotor in motion after the kick, which is $t_{relaxation}$ divided by the total time $T_+$ the rotor is in motion in the clockwise direction. This leads to}
%This probability is the time d$t$ between these two velocities over the total time the rotor needs to relax, $t_\textrm{relaxation}$,

%\begin{equation}
%  h\left(\Omega|\Omega_{0}\right)\textrm{d}\Omega=\frac{\mathrm{d}t}{t_\textrm{relaxation}}=\frac{1}{t_\textrm{relaxation}\left|\frac{\mathrm{d}\Omega}{\mathrm{d}t}\right|}\,\mathrm{d}\Omega=\frac{C}{|\dot{\Omega}|}\,\mathrm{d}\Omega\, ,\qquad C=\frac{1}{t_\textrm{relaxation}}
%\end{equation}

\begin{equation}
h(\Omega | \Omega_0) d\Omega = \frac{dt}{T_\pm} =  \frac{1}{T_\pm |\frac{d\Omega}{dt}|} = \frac{C_\pm}{|\dot{\Omega}|}\,,\quad C_\pm = \frac{1}{T_\pm}
\end{equation}\label{eq:FRelax1}

$\fl\hspace{3.8cc}$or more precisely,

\begin{equation}
h\left(\Omega|\Omega_{0}\right)=\begin{cases}
\frac{C_\pm}{\left|\dot{{\Omega}}\right|} & \mathrm{\: when}\:\sfrac{\Omega_{0}}{\Omega}>1\\
0 & \:\mathrm{when}\:\sfrac{\Omega_0}{\Omega}\leq1
\end{cases}\label{eq:FRelax2}
\end{equation} 

$\fl\hspace{3.8cc}$because the magnitude of the angular velocity $\Omega$ has to be lower than that of the initial angular velocity $\Omega_0$. In addition, $\Omega$ and $\Omega_0$ must have the same sign.

Finally, with these two ingredients we can obtain define the regular part of the AVD as the multiplication between the probabilities $h\left(\Omega|\Omega_{0}\right)$ and $G\left(\Omega_{0}\right)$, and integrating over all possible initial velocities $\Omega_0$,

\begin{equation}
f_{R}\left(\Omega\right)  =  \int_{-\infty}^{\infty}\textrm{d}\Omega_{0}\,h\left(\Omega|\Omega_{0}\right)\,G\left(\Omega_{0}\right)\,\Theta\left(\Omega_{0}-\Omega\right)
\end{equation}
\begin{equation}
f_{R}\left(\Omega\right)=\rho L\sqrt{\frac{T}{2\pi m}}\left[\Theta\left(-\Omega\right)\frac{C_{-}}{\left|\dot{\Omega}\right|}\int_{0}^{1}\textrm{d}z\,\textrm{e}^{-\frac{\left(1+\xi z^{2}\right)^{2}I}{2T\left(1+\alpha_{-}\right)^{2}\xi z^{2}}\Omega^{2}}+\Theta\left(\Omega\right)\frac{C_{+}}{\left|\dot{\Omega}\right|}\int_{0}^{1}\textrm{d}z\,\textrm{e}^{-\frac{\left(1+\xi z^{2}\right)^{2}I}{2T\left(1+\alpha_{+}\right)^{2}\xi z^{2}}\Omega^{2}}\right]
\label{eq: AVD SKR}
\end{equation}

$\fl\hspace{3.8cc}$where $C_\pm$ are normalisation constants and $\Theta$ is the Heaviside step-function ($\Theta(x)=1$ for $x>0$ and 0 otherwise). {Note that when the symmetry is broken (i.e., $\alpha_+ \neq \alpha_-$) there is a difference between $C_+$ and $C_-$. This is due to the fact that although the kick events are distributed symmetrically for the clockwise and anti-clockwise directions, their magnitudes are not and therefore also the total amount of time that the rotor is in motion is different for the clockwise and anti-clockwise directions.}

This expression is almost identical to the one obtained by Talbot \textit{et al}. \cite{talbot_kinetics_2011}, with the small but significant modification that allows for a non-constant acceleration, $\dot{\Omega}(\Omega)$. This acceleration was obtained experimentally, and fitted to the function 
\begin{equation}
  \dot{\Omega}\,(\Omega)\,=\,A\,\textrm{asinh}(B\,\Omega) \label{eq:acceleration}
\end{equation}  

$\fl\hspace{3.8cc}$for $A$ and $B$ constants, as was discussed above (cf. Fig.~\ref{fig:AvsV}-b). 

Therefore, once this acceleration function is considered in the model, the only free parameter is the (slightly modified) granular temperature, $(1+\alpha_\pm)^2\,T$. 

 \begin{figure}[h!]
  \begin{center}
   \includegraphics[scale=0.30]{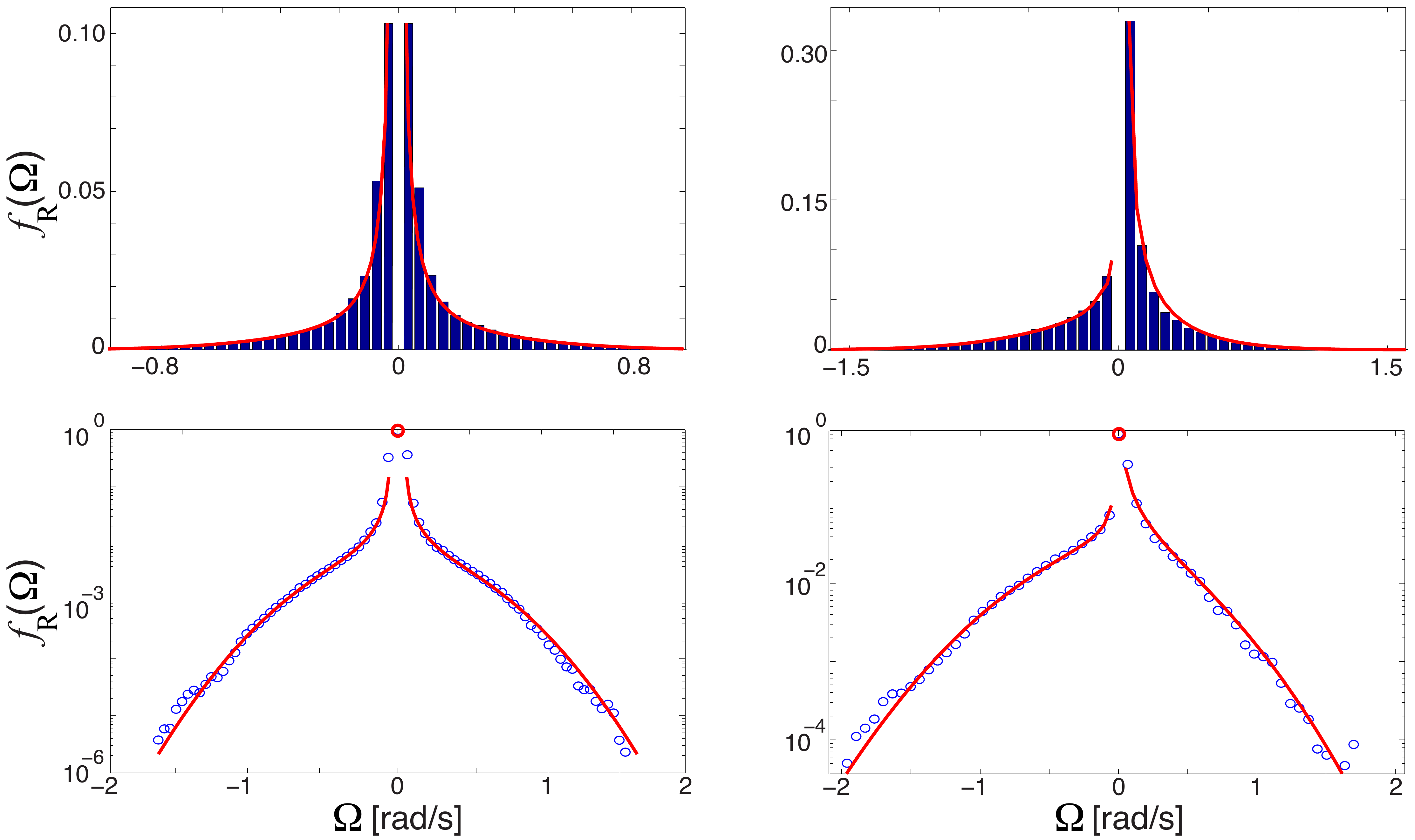}  
  \end{center}
  \caption{Regular part of the AVD in the Single Kick Limit for the symmetric (left) and non-symmetric (right) cases; the vertical axes on the top and bottom panels are linear and logarithmic respectively. Blue bars and circles correspond to the experimental data (for $f=20$ Hz and $a=1.4$ mm) and the red line represents the model fitted to the experimental data, using the granular temperature as a free parameter.}
  \label{fig:AVDKickRegime}
  \end{figure}

To verify the model, we conducted experiments for different granular temperatures, keeping the density of the gas and the amplitude of the shaker constant while changing the frequency, $f=[20,\,22,\,24,\,26,\,28] $ Hz. Measurements were taken for several hours for each frequency in order to ensure proper statistics. The AVD for two representative measurements are shown in Fig.~\ref{fig:AVDKickRegime} (blue circles and bars), both in linear (upper plots) and semi-logarithmic (lower plots) scale. For every frequency, we measured with a symmetric (left plots) and non-symmetric (right plots) rotor, where in the latter case the symmetry is broken by a neoprene sealing strip which is mounted on the right side of each vane. This asymmetry is observed in the data where it is not only seen that the rotor reaches larger angular velocities in the anti-clockwise than the clockwise direction, but also that the shape of the AVD curve is different on each side because the coefficient of restitution is different; this dependence is in accordance with the model described earlier (Eq. \ref{eq: AVD SKR}). {It is important to note that the free parameter $(1+\alpha_+)^2T$ in the symmetric and the asymmetric cases (i.e. in the clockwise direction), for which the restitution coefficient $\alpha_+$ is the same due to the fact that this side of each vane remains uncoated, are in good agreement.}

We can obtain the singularity of the AVD through the relation in Eq. \ref{eq:AVD}. This singularity is plotted as a red dot in  
Fig.~\ref{fig:AVDKickRegime} (lower plots). In the linear scale, just the regular part is plotted and the singularity is omitted explaining the empty space around zero.    

Clearly, the model excellently describes the experimental observations. Our modified model is thus capable of describing the behaviour of the rotor in the Single Kick Limit, including velocities close to zero unlike previous models. We believe that it is also very likely to describe the discrepancies found in the study by Gnoli \textit{et al} \cite{gnoli_granular_2013} between their experiments and the theory of Talbot \textit{et al} \cite{talbot_kinetics_2011}, but we do not have access to their angular acceleration function $\dot{\Omega}(\Omega)$ to probe it. To study the rotor behaviour beyond the Single Kick Limit we extend our analysis in the next Section \ref{s:B-SKL} by relaxing the previous assumption that the rotor always reaches a state of rest after a kick.

%%%%%%%%%%%%%%%%%%%%%%%%%%%%%%%%%%%%%%%%%%%%%%%%%%%%%%%%%%%%%%%%%%%%%%%%%%%%%%%%%%%%%%%%%%%%%%%%%%%%%%%%%%%%%%%%%%%%%%
 
 \section{Beyond the Single Kick Limit}\label{s:B-SKL}

With the goal of understanding the behaviour of the granular rotor beyond the Single Kick Limit, we inject more energy to the granular gas and, as a result, the time between particle-vane collisions becomes comparable to the time the rotor needs to be stopped by friction ($\tau_{c}\sim\tau_{s}$). In this regime, the model developed in the previous section can not describe the angular velocity distribution because now the rotor does not reach the rest position after every collision. However, we can extend the previous model by considering the additional probability that a new collision occurs while the rotor is still in motion.      

 \begin{figure}[h!]
  \begin{center}
  \includegraphics[scale=0.5]{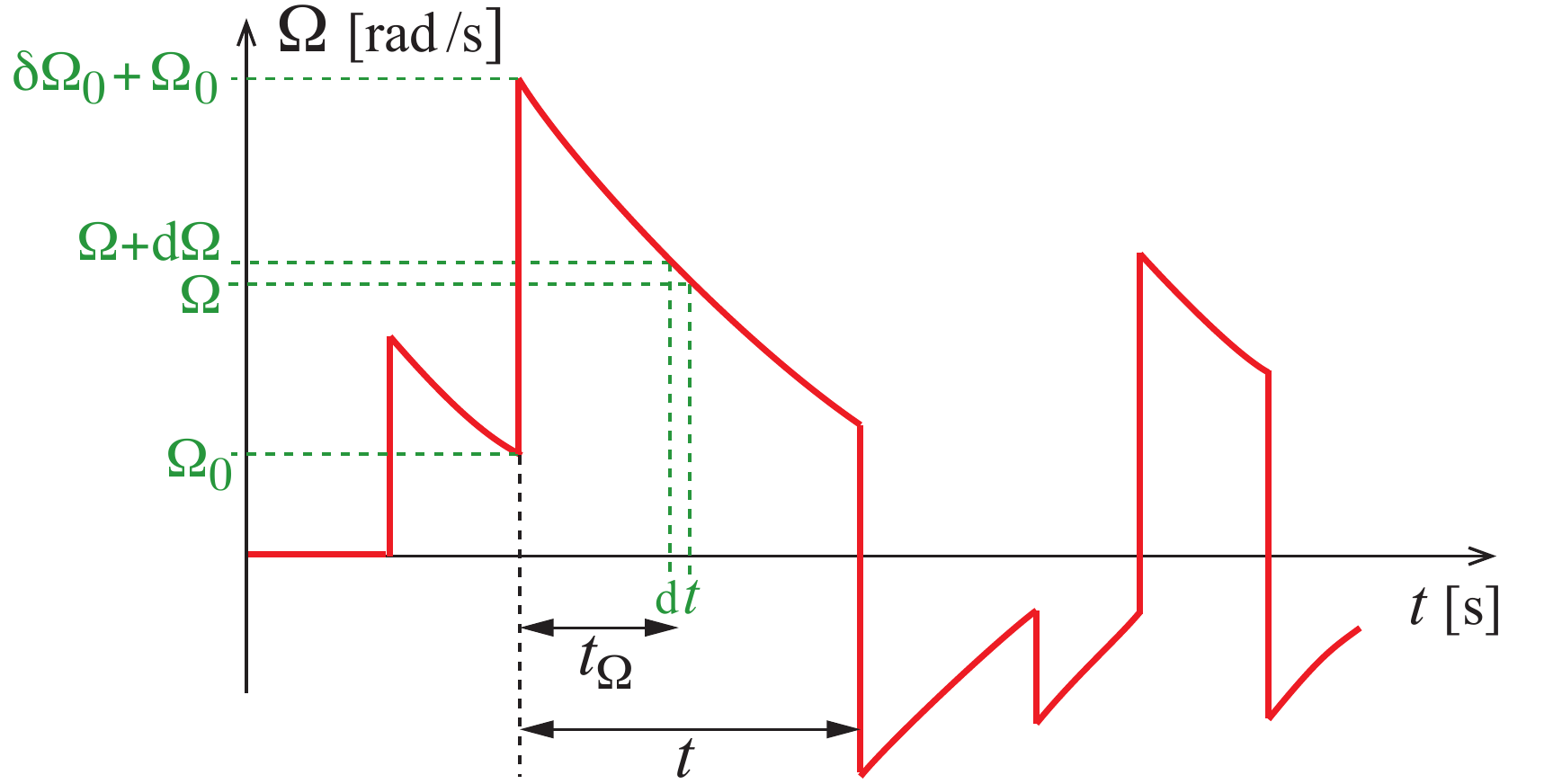}
 \end{center}
\caption{Schematic of the time evolution of the angular velocity beyond the Single Kick Limit. When the rotor is turning with velocity $\Omega_0$, a kick with initial velocity $\delta\Omega_0$ occurs, and the resulting initial velocity is the sum of both, i.e., $\delta\Omega_0+\Omega_0$. Then, the time that the rotor takes to reach a velocity $\Omega$ will be $t_\Omega$ provided that no subsequent kick occurs before this time.}\label{fig:schematicBeyondKickRegime}
 \end{figure}

When the rotor is rotating with velocity $\Omega_0$ and a kick with initial velocity $\delta\Omega_0$ occurs,  the resulting initial velocity after the kick will be the sum of both, $\delta\Omega_0+\Omega_0$, as shown in the Fig.~\ref{fig:schematicBeyondKickRegime}. Subsequently, we define $t_\Omega$ as the time that the rotor takes to reach a velocity $\Omega$ after the kick.

To describe the AVD we must therefore consider the kick distribution $G(\delta\Omega_0)$ and the probability to subsequently find the rotor at a certain velocity $\Omega$, just as in the Single Kick Limit. But a third ingredient has to be considered, which is the distribution of the time between two kicks which we will denote as $g(t)$. This is because the probability to find the rotor with a velocity $\Omega$ starting from an initial velocity $\Omega_0+\delta\Omega_0$ does not depend only on $h\propto\frac{1}{\left|\dot{\Omega}\right|}$  (as Eq. \ref{eq:FRelax2}), but also on the probability to not have a subsequent second kick before the time $t_\Omega$.

\begin{figure}[h!]
  \begin{center}
  \includegraphics[scale=0.38]{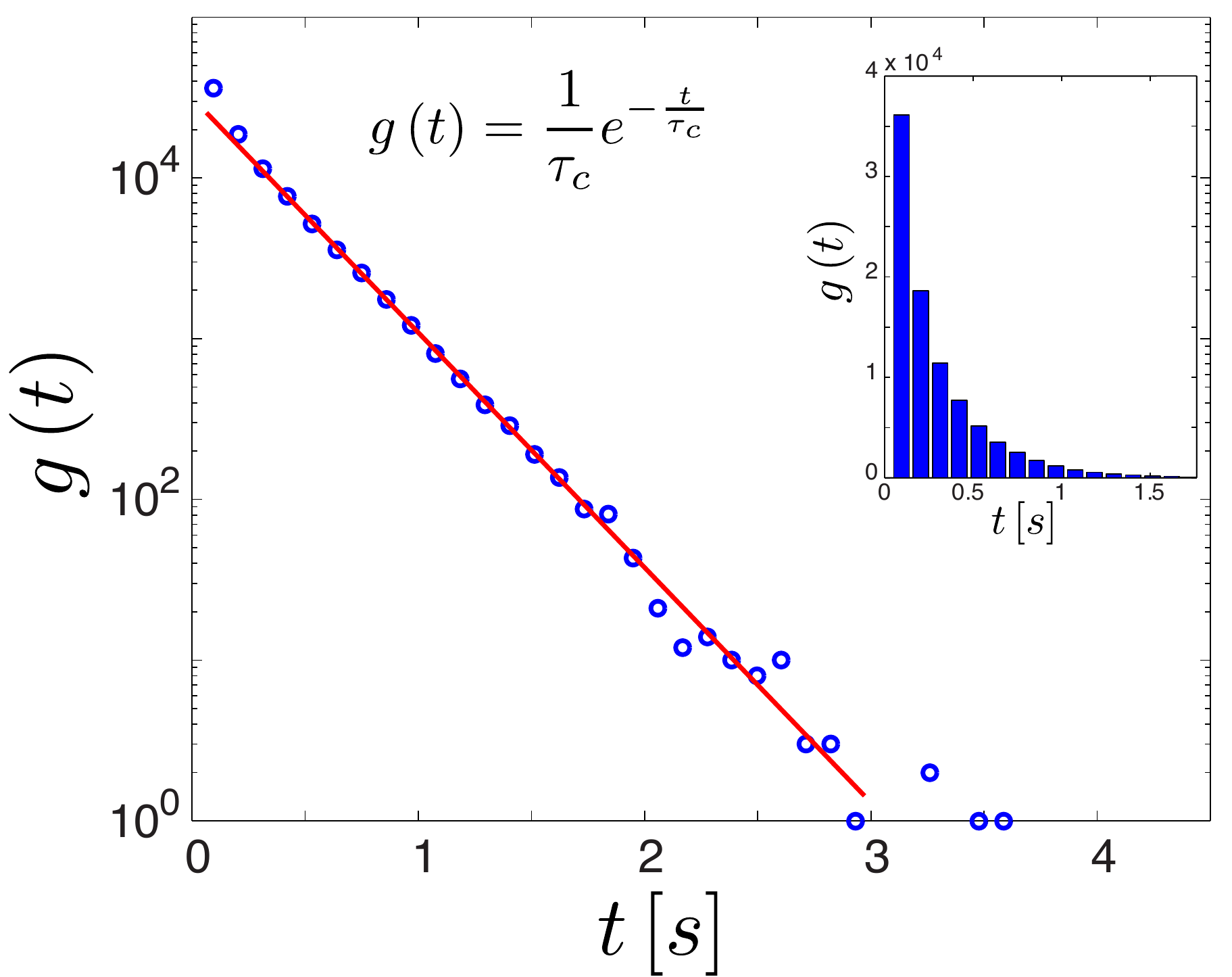}
  \end{center}
  \caption{Experimentally determined distribution of the time interval between two subsequent kicks (for $f=20$ Hz, $a=1.4$ mm). This distribution is exponential 
  and depends only on the average time between particle-vane collisions $\tau_c$}
  \label{fig:DistrTimeKick}
 \end{figure}

For a diluted granular gas the particle collisions are uncorrelated and the particle-vane collisions are expected to be described by a Poisson process \cite{brilliantov_kinetic_2004}. Therefore, we expect that the distribution of the time between kicks will have an exponential shape, i.e., $g\left(t\right)=\frac{1}{\tau_c}e^{-\frac{t}{\tau_c}}$, which depends only on the average time between particle-vane collisions $\tau_c$. This distribution is obtained and corroborated experimentally, as shown in Fig.~\ref{fig:DistrTimeKick}.

While the behaviour of the probability $h$ with respect to the angular acceleration $\dot{\Omega}$ is the same as that in the Single Kick Limit, the final expression is in a subtle manner different to that obtained in Eq. \ref{eq:FRelax2}. From Fig.~\ref{fig:schematicBeyondKickRegime}, this probability is observed to be the time $\textrm{d}t$ between $\Omega$ and $\Omega+\textrm{d}\Omega$ over the total time until the next collision $t$; the important difference from the previous expression is that this time is not constant, and therefore not merely dependent on the initial velocity $\Omega_0+\delta\Omega_0$

\begin{equation}
  h\left(\Omega|\Omega_{0}+\delta\Omega_0\right)\textrm{d}\Omega=\frac{\mathrm{d}t}{t}=\frac{1}{t\,\left|\frac{\mathrm{d}\Omega}{\mathrm{d}t}\right|}\,\mathrm{d}\Omega=\frac{1}{|\dot{\Omega} | \, t}\,\mathrm{d}\Omega
\end{equation}\label{eq:FRelax3}

$\fl\hspace{3.8cc}$or more precisely,
\begin{equation}
h\left(\Omega|\Omega_{0}+\delta\Omega_0\right)=\begin{cases}
\left|\frac{1}{\dot{\Omega}\, t}\right| & \mathrm{\: when\:\sfrac{\left(\Omega_{0}+\delta\Omega_0\right)}{\Omega} >1}\\
0 & \:\mathrm{when}\:\sfrac{\left(\Omega_{0}+\delta\Omega_0\right)}{\Omega} \leq 1
\end{cases}\label{eq:FRelax4}
\end{equation} 

Hence, in this regime the probability to find the rotor with angular velocity between $\Omega$ and $\Omega+\delta\Omega$ is found by taking the product of the collision time distribution $g(t)$ and the probability $h\left(\Omega|\Omega_{0}+\delta\Omega_0\right)$ and subsequently integrating this product all collision times occurring after $t_\Omega$ (which corresponds to the probability that the next collision will happen only \emph{after} the velocity $\Omega$ is reached),

\begin{eqnarray}
 \hspace{-1cm}H\left( \Omega|\Omega_{0}+\delta\Omega_0\right) &=& \Theta\left(\left| 
 \delta\Omega_0+\Omega_0\right|-\left|\Omega\right|\right)\,\int_{t_\Omega}^\infty g(t)\, 
  h\left(\Omega|\delta\Omega_0+\Omega_0\right)\,\textrm{d}t\\
  &=& -\left|\frac{1}{\dot{\Omega}\tau_c}\right| \Theta\left(|\delta\Omega_{0}+\Omega_{0}|-|\Omega|\right)\, \textrm{ei}\left(-\frac{t_\Omega}{\tau_c}\right)
\end{eqnarray}

$\fl\hspace{3.8cc}$where $\textrm{ei}\,(x)=\int_{-\infty}^x\frac{1}{t}\,\textrm{e}^t\textrm{d}t$ is the exponential integral function. Now finally, the probability $f(\Omega)$d$\Omega$ is determined by multiplying the probability $H\left( \Omega|\Omega_{0}+\delta\Omega_0\right)$d$\Omega$ of finding $\Omega$ from the initial value $\Omega_0+\delta\Omega_0$ with the probability $G(\delta\Omega_0)$d$\delta\Omega_0$ of having a kick of size $\delta\Omega_0$ and with the probability of having an initial angular velocity $\Omega_0$, i.e., with $f(\Omega_0)$d$\Omega_0$. Subsequently this products needs to be integrated over both $\delta\Omega_0$ and $\Omega_0$, leading to 

\begin{equation}
  f(\Omega)=\int_{-\infty}^{\infty}\textrm{d}\Omega_0\int_{-\infty}^\infty\textrm{d}\delta\Omega_0\,G(\delta\Omega_0)\,H(\Omega|\Omega_{0}+\delta\Omega_0)\,f(\Omega_0)
\label{eq:AVDBeyond1}
\end{equation}

This equation contains the AVD $f(\Omega)$ both on the left and in the integrand on the right and therefore constitutes an integral equation, which does not have an analytical solution. To solve it numerically, we first define the kernel function 
$K(\Omega,\Omega_0)=\int_{-\infty}^\infty\textrm{d}\delta\Omega_0\,G(\delta\Omega_0)\,H(\Omega|\Omega_{0}+\delta\Omega_0)$ and rewrite Eq. \ref{eq:AVDBeyond1} as
\begin{equation}
  f(\Omega)=\int_{-\infty}^{\infty}\textrm{d}\Omega_0\,K(\Omega,\Omega_0)\,f(\Omega_0)\label{eq:AVDBeyond2}
\end{equation}

{This integral equation is known as a homogeneous Fredholm equation of the second type. One of the standard methods of solving such an equation is through discretisation of the integral which then directly leads to a matrix eigenvalue problem, which is the approach that we take in the following \cite{atkinson_numerical_2009}. Different discretisation schemes are used for the symmetric and non-symmetric cases in Eq. \ref{eq:AVDBeyond2}.} For the symmetric case, we can separate the integral in Eq. \ref{eq:AVDBeyond2} in two parts: positive and negative angular velocities. 

\begin{equation}
  f\left(\Omega\right)  = \int_{-\infty}^{0}\textrm{d}\Omega_{0}\, K\left(\Omega,\Omega_{0}\right)f\left(\Omega_{0}\right)+\int_{0}^{\infty}\textrm{d}\Omega_{0}\, K\left(\Omega,\Omega_{0}\right)f\left(\Omega_{0}\right)
\end{equation}

With the variable change $\Omega_0=-\Omega_0$ in the first integral of the distribution and using that in the symmetric case the AVD follows $f(\Omega_0)=f(-\Omega_0)$, the distribution can be written as,
\begin{eqnarray}
f\left(\Omega\right) & = & \int_{0}^{\infty}\textrm{d}\Omega_{0}\left(K\left(\Omega,-\Omega_{0}\right)+K\left(\Omega,\Omega_{0}\right)\right)f\left(\Omega_{0}\right)\nonumber \\
 & = & \int_{0}^{\infty}\textrm{d}\Omega_{0}\,\widetilde{K}\left(\Omega,\Omega_{0}\right)f\left(\Omega_{0}\right)\label{eq:AVDBeyond3}
\end{eqnarray}

Now we can discretize the equation, restricting ourselves to an array of nonnegative, equidistant ($\Delta\Omega$) values $\Omega_i$
\begin{equation}
 f(\Omega_i)=\sum_{j=1} ^{N} \widetilde{K}(\Omega_i,\Omega_j)\,f(\Omega_j)\,\Delta\Omega
\end{equation}

$\fl\hspace{3.8cc}$and rewrite it in matrix notation

\begin{equation}
 \vec{f}=\widetilde{\textrm{\textbf{K}}}\,\vec{f}\quad,\,\widetilde{\textrm{\textbf{K}}}=[\,\widetilde{K}^{i,j}\,]\Delta\Omega=[\,\widetilde{K}(\Omega_i,\Omega_j)\,]\Delta\Omega  \label{eq:AVDBeyond4}
\end{equation}\\
In this way, the expression for the AVD (Eq. \ref{eq:AVDBeyond2}) was converted to a matrix eigenvalue problem, and the 
eigenvector of $\textbf{K}$ with eigenvalue equal to one corresponds to the (approximate) solution of AVD. 

This expression Eq. \ref{eq:AVDBeyond4}, holds only when the rotor is symmetric, and when the rotor is asymmetric we have to include additional conditions related to the direction of the kicks, which will now produce different angular momentum changes. 
This is treated in detail in \ref{ap:BSKR} and results in the following equation for the AVD
\begin{eqnarray}
 \hspace{-1cm}f_{-}\left(\Omega\right) &=& \int_{-\infty}^{0}\mathrm{d}\Omega_{0}\, K_{1}\left(\Omega,\Omega_{0}\right)f_{-}\left(\Omega_{0}\right)+\int_{0}^{\infty}\mathrm{d}\Omega_{0}\, K_{2}\left(\Omega,\Omega_{0}\right)f_{+}\left(\Omega_{0}\right)\nonumber\\
 \hspace{-1cm}f_{+}\left(\Omega\right) &=& \int_{-\infty}^{0}\mathrm{d}\Omega_{0}\, K_{3}\left(\Omega,\Omega_{0}\right)f_{-}\left(\Omega_{0}\right)+\int_{0}^{\infty}\mathrm{d}\Omega_{0}\, K_{4}\left(\Omega,\Omega_{0}\right)f_{+}\left(\Omega_{0}\right)
\label{eq:NSSysEq}\end{eqnarray}

Here the subscript symbol $\pm$ with each function indicates if the angular velocity inside its argument is positive or negative. Similar to the symmetric case (Eq. \ref{eq:AVDBeyond2} and \ref{eq:AVDBeyond3}), we can now write and solve the system of integral equations (Eq. \ref{eq:NSSysEq}) as a matrix eigenvalue problem (see \ref{ap:BSKR}),

\begin{equation}
\vec{f}=\mathrm{\mathbf{K}}\cdot\vec{f}\qquad,\,\mathrm{\mathbf{K}}=\left[\begin{array}{cc}
K_{1}^{i,j} & K_{2}^{i,j}\\
K_{3}^{i,j} & K_{4}^{i,j}
\end{array}\right]\Delta\Omega
\end{equation}

\begin{figure}[h!]
  \begin{center}
   \includegraphics[scale=0.30]{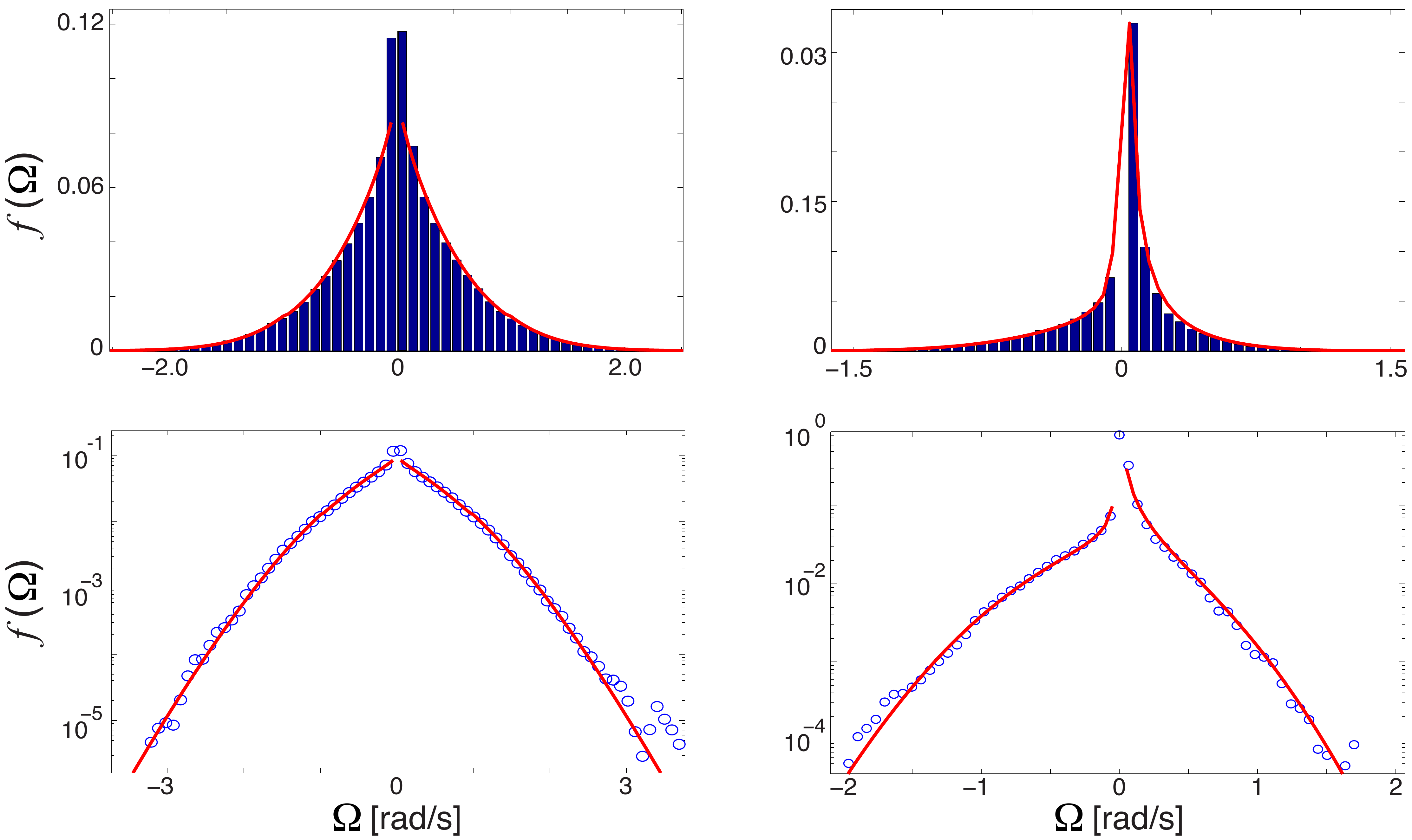}  
  \end{center}
  \caption{AVD beyond the Single Kick Limit for the symmetric (left) and non-symmetric (right) cases; the vertical axes on the top and bottom panels are linear and logarithmic respectively. Blue bars and circles correspond to the experimental data (for $f=30$ Hz and $a=1.4$ mm) and the red line represents the model fitted to the experimental data, using the granular temperature as a free parameter.}
  \label{fig:AVDBeyondKick}
  \end{figure}

To verify the extended model, we conducted experiments for different granular temperatures, keeping the density of the gas 
and the amplitude of the shaker constant while changing the frequency, $f=\,30,\,32,\,34$ and 36 Hz. Figure \ref{fig:AVDBeyondKick} shows a representative measurement, $f=30$ Hz,  (blue circles and bars), in linear (upper plots) and semi-logarithmic (bottom plots) scale. For every frequency, we measure with a symmetric (left plots) and a non-symmetric (right plots) rotor. 

The angular acceleration $\dot{\Omega}$ is required in the model and is obtained in the same way as in the Single Kick Limit, (refer to section \ref{s:SKL} and Fig.~\ref{fig:AvsV}-b); and again the granular temperature ($T$) is the only free parameter. Fig.~\ref{fig:AVDBeyondKick} shows that the solution of the model has a very good agreement with the experimental data. Therefore this model is capable of describing the AVD of the rotor beyond the Single Kick Limit and is the first model to achieve this. Of course, also the Single Kick Limit and the Brownian Limit are expected to be included in this model for the intermediate regime. Because the model depends on the collision time ($\tau_c$), the rotor is in the Single Kick Limit when $\tau_c$ tends towards infinity when compared to the relaxation time $\tau_s$. In other words, in this limit a collision practically never occurs when the rotor is in movement. When on the other hand $\tau_c$ tends to zero this results in frequent collisions and a negligible change of the angular velocity of the rotor in between kicks, such that it is in the Brownian Limit.

\begin{figure}[h!]
 \begin{center}
 \hspace{-0.25cm} 
 \includegraphics[scale=0.42]{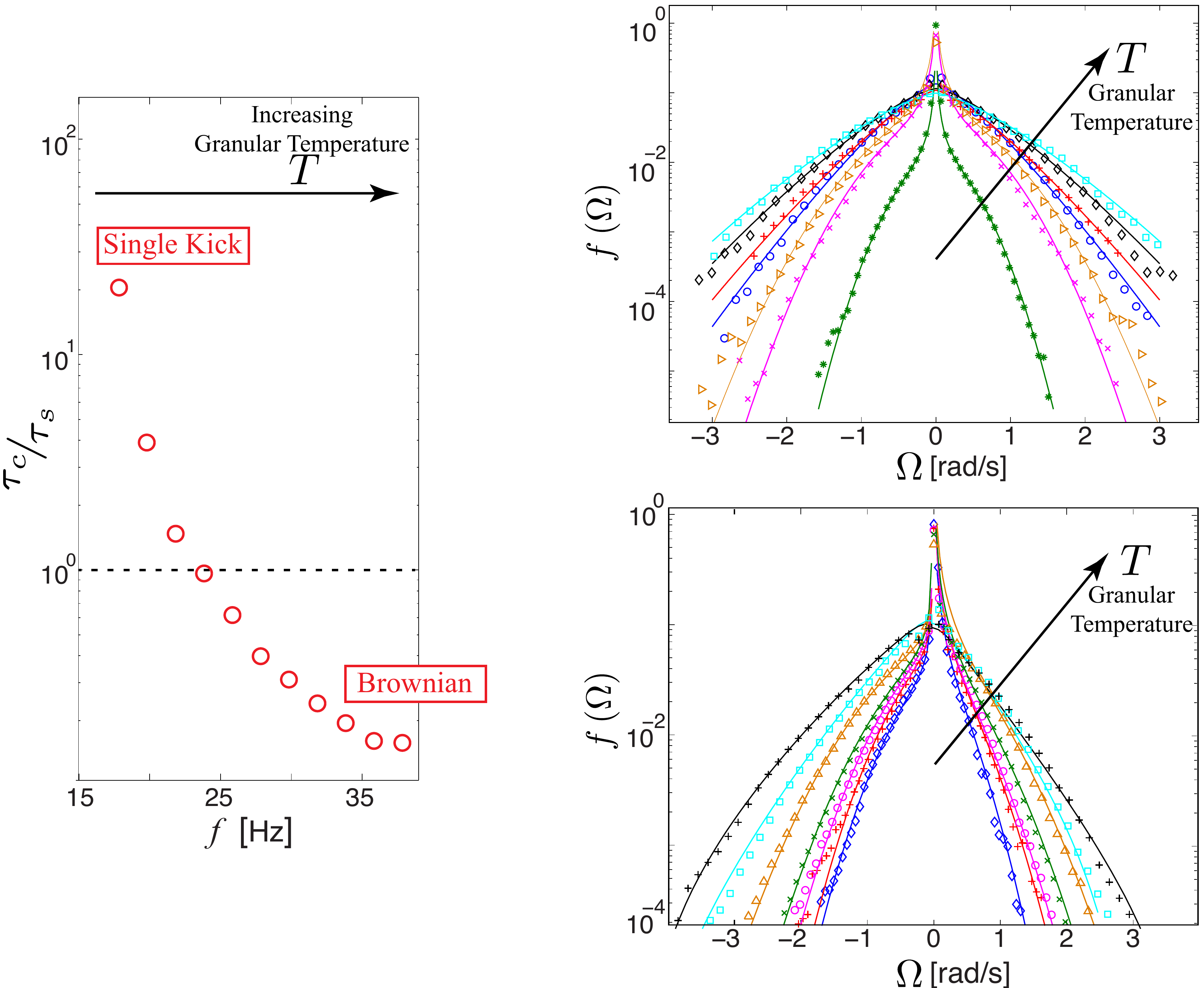}
\end{center}
 \caption{The AVD $f(\Omega)$ is plotted for increasing granular temperatures, for both the symmetric (right-top) and the non-symmetric (right-bottom) case. The symbols correspond to the experimental data (for $f=20$, 22, 24, 26, 28, 30 and 32 Hz and $a=1.4$ mm) whereas the lines represent the extended model. Clearly there is a very good agreement between model and experiment. On the left, the ratio of collision time to relaxation time ($\tau_c/\tau_s$) versus frequency is plotted, showing how the system is moving from the Single Kick Limit to the Brownian Limit.}
 \label{fig:AVDALLKickRegime}
\end{figure}

In order to demonstrate this, we increase the granular temperature by increasing the frequency of the shaker ($f=20$ Hz$\rightarrow36$ Hz), such that the system starts in the Single Kick Limit and moves towards the Brownian Limit, such as is shown in Fig.~\ref{fig:AVDALLKickRegime}-left. In this figure we plot the ratio of the collision time to the relaxation time ($\tau_c/\tau_s$) versus frequency, showing that the system is moving from high to low ratio values, and therefore from the Single Kick to the Brownian Limit. In Fig.~\ref{fig:AVDALLKickRegime} (right) we show the experimental data together with the extended model results, corroborating the accuracy of the extended model for the AVD of a granular rotor throughout the explored parameter space and not just in limiting conditions.  

%%%%%%%%%%%%%%%%%%%%%%%%%%%%%%%%%%%%%%%%%%%%%%%%%%%%%%%%%%%%%%%%%%%%%%%%%%%%%%%%%%%%%%%%%%%%%%%%%%%%%%%%%%%%%%%%%%%%%%

%%%%%%%%%%%%%%%%%%%%%%%%%%%%%%%%%%%%%%%%%%%%%%%%%%%%%%%%%%%%%%%%%%%%%%%%%%%%%%%%%%%%%%%%%%%%%%%%%%%%%%%%%%%%%%%%%%%%%%

\section{Conclusions}\label{s:Conclusions}

We built a rotor composed of four vanes, similar to the Smoluchowski-Feynman device, immersed in a granular gas, to experimentally study the behaviour of its angular velocity distribution (AVD) under the influence of the unavoidable external friction that is present in the ball bearing connecting the rotor to the container. The rotor can be turned into a ratchet by breaking the symmetry, which is achieved by covering one side of each vane of the rotor with a neoprene strip. The granular gas is created by a piston moving in our container that is connected to an electromagnetic shaker. In our experiments we fix the number of particles and modify the properties of the granular gas by changing the frequency of the driving, i.e., we change the granular temperature of the gas by varying the amount of injected energy.
 
We first analyse the AVD of the rotor for low temperatures of the granular gas, i.e., in the Single Kick Limit, when the rotor is in rest for most of the time. We show that the external friction plays an important role in the rotor behaviour and quantify its effect by measuring the angular acceleration, which we observe to be non-constant as a function of the angular velocity. Closely following the model studied by Talbot \textit{et al.} \cite{talbot_kinetics_2011}, we develop a model incorporating the details of the angular acceleration that we show to be capable to describe the AVD and obtain a good agreement between the model and experimental results.

With the aim to describe the AVD of the rotor in the entire parameter space, we increase the temperature of the granular gas in order to experimentally push the system beyond the Single Kick Limit. We extend our model into a new one that can describe the AVD not only for this intermediate condition, but also in its limiting behaviours. Finally, we show that this extended model agrees very well with the experimental data.   

%%%%%%%%%%%%%%%%%%%%%%%%%%%%%%%%%%%%%%%%%%%%%%%%%%%%%%%%%%%%%%%%%%%%%%%%%%%%%%%%%%%%%%%%%%%%%%%%%%%%%%%%%%%%%%%%%%%%%%

%%%%%%%%%%%%%%%%%%%%%%%%%%%%%%%%%%%%%%%%%%%%%%%%%%%%%%%%%%%%%%%%%%%%%%%%%%%%%%%%%%%%%%%%%%%%%%%%%%%%%%%%%%%%%%%%%%%%%%

\appendix

\section{Description of the angular velocity distribution model beyond Single Kick Limit} \label{ap:BSKR}

In this appendix, we will discuss the derivation of the model beyond the Single Kick limit, as presented in Section \ref{s:B-SKL}, in greater depth.  

When the rotor is rotating with velocity $\Omega_0$ and a kick with initial velocity $\delta\Omega_0$ occurs,  the initial velocity will the sum of both $\delta\Omega_0+\Omega_0$, as shown in Fig.~\ref{fig:schematicBeyondKickRegime}. 

To describe the AVD, we first need to find the probability of the rotor having an angular velocity between $\Omega$ and $\Omega+\textrm{d}\Omega$. We define $t_\Omega$ as the time the rotor takes to reach the velocity $\Omega$ after a kick. This time is obtained from the acceleration function

\begin{equation}
  \dot{\Omega}=A\,\textrm{asinh}(B\,\Omega)
\end{equation} 

$\fl\hspace{3.8cc}$and the initial condition $\Omega(t=0)=\delta\Omega_0+\Omega_0$. Solving this equation for $\Omega(t)$ and determining the time that the rotor takes to reach a angular velocity $\Omega$ leads to
\begin{equation}
t_\Omega=\frac{1}{AB}\,\left(\textrm{Chi}(\left|\textrm{asinh}(B\Omega)\right|)-\textrm{Chi}(\left|\textrm{asinh}(B(\delta\Omega_0+\Omega_0))\right|)\right)\quad,\,A<0
\end{equation}

$\fl\hspace{3.8cc}$where Chi($\Omega$) is the hyperbolic cosine integral. Considering the situation described above, the rotor will not reach velocity $\Omega$ if a new collision occurs within this time interval $t_\Omega$ and thus it is important to consider the distribution of the time intervals between consecutive kicks. This distribution is obtained experimentally and has a exponential shape, as was shown in Fig.~\ref{fig:DistrTimeKick},

\begin{equation}
  g(t)=\frac{1}{\tau_c}e^{-^t/_{\tau_c}}
\end{equation}

$\fl\hspace{3.8cc}$where $\tau_c$ is the average time interval between particle-vane collisions. Therefore, when the next collision occurs at a time $t>t_\Omega$, the probability to find the rotor at a certain angular velocity between $\Omega$ and $\Omega+\mathrm{d}\Omega$ after a kick with initial angular velocity $\delta\Omega_0+\Omega_0$ will be the corresponding infinitesimal time interval $\textrm{d}t$ divided by the entire time interval until the next collision $t$, and we can write 

\begin{equation}
  h\left(\Omega|\Omega_{0}+\delta\Omega_0\right)\textrm{d}\Omega=\frac{\mathrm{d}t}{t}=\frac{1}{t\,\frac{\mathrm{d}\Omega}{\mathrm{d}t}}\,\mathrm{d}\Omega=\frac{1}{\dot{\Omega} \, t}\,\mathrm{d}\Omega
\end{equation}

Imposing the condition that the magnitude of the angular velocity of the rotor has to be always lower than the magnitude of the initial angular velocity, this probability density is

\begin{equation}
  h\left(\Omega|\Omega_{0}+\delta\Omega_0\right)\textrm{d}\Omega=\frac{\mathrm{d}t}{t}=\frac{1}{t\,\left|\frac{\mathrm{d}\Omega}{\mathrm{d}t}\right|}\,\mathrm{d}\Omega=\frac{1}{|\dot{\Omega} | \, t}\,\mathrm{d}\Omega
\end{equation}\label{eq:frelax-1}

Hence, considering the probability $h(\Omega|\delta\Omega_0+\Omega_0)\textrm{d}\Omega$ and the distribution of time between kicks $g(t)$, the probability of the rotor having a angular velocity between $\Omega$ and $\Omega+\textrm{d}\Omega$ at any time is the multiplication of both functions and integrated over all times larger than $t_\Omega$, which leads to

\begin{eqnarray}
\hspace{-1cm}H_{-}\left(\Omega|\Omega_{0}+\delta\Omega_0\right) & = & \Theta\left(\Omega-\left(\delta\Omega_{0}+\Omega_{0}\right)\right)\int_{t_\Omega}^{\infty}g\left(t\right)\cdot h\left(\Omega|\Omega_{0}+\delta\Omega_0\right)\,\textrm{d}t\\
\hspace{-1cm}H_{+}\left(\Omega|\Omega_{0}+\delta\Omega_0\right) & = & \Theta\left(\left(\delta\Omega_{0}+\Omega_{0}\right)-\Omega\right)\int_{t_\Omega}^{\infty}g\left(t\right)\cdot h\left(\Omega|\Omega_{0}+\delta\Omega_0\right)\,\textrm{d}t
\end{eqnarray}

Here, $H_-$ and $H_+$ refers to the rotor moving in anti-clockwise (negative) and clockwise (positive) direction respectively. Using the exponential integral definition, $\textrm{ei}\left(x\right)=\int_{-\infty}^{x}\frac{1}{t}\,\textrm{e}^{t}\textrm{d}t$, this distribution can be solved analytically:

\begin{eqnarray}
\hspace{-1cm}H_{-}\left(\Omega|\Omega_{0}+\delta\Omega_0\right) & = & \Theta\left(\Omega-(\delta\Omega_{0}+\Omega_{0})\right)\int_{t_\Omega}^{\infty}\frac{1}{\tau_c}\,\textrm{e}^{-^{t}/_{\tau_c}}\left|\frac{1}{\dot{\Omega}t}\right| \,\textrm{d}t\nonumber \\
 & = & -\left|\frac{1}{\tau_c\,\dot{\Omega}}\right| \Theta\left(\Omega-(\delta\Omega_{0}+\Omega_{0})\right)\,\textrm{ei}\left(-\frac{t_\Omega}{\tau}\right)\\
\hspace{-1cm}H_{+}\left(\Omega|\Omega_{0}+\delta\Omega_0\right) & = & \Theta\left((\delta\Omega_{0}+\Omega_{0})-\Omega\right)\int_{t_\Omega}^{\infty}\frac{1}{\tau_c}\,\textrm{e}^{-^{t}/_{\tau_c}}\left|\frac{1}{\dot{\Omega}t}\right|\,\textrm{d}t\nonumber \\
 & = & -\left|\frac{1}{\tau_c\,\dot{\Omega}}\right|\Theta\left((\delta\Omega_{0}+\Omega_{0})-\Omega\right)\,\textrm{ei}\left(-\frac{t_\Omega}{\tau}\right)
\end{eqnarray}

Therefore, the AVD $f(\Omega)$ is determined by multiplying the probability $H\left( \Omega|\Omega_{0}+\delta\Omega_0\right)$d$\Omega$ of finding $\Omega$ from the initial value $\Omega_0+\delta\Omega_0$ with the probability $G(\delta\Omega_0)$d$\delta\Omega_0$ (Eq. \ref{eq:KickDistribution}) of having a kick of size $\delta\Omega_0$ and with the probability $f(\Omega_0)$d$\Omega_0$ of having an initial angular velocity $\Omega_0$. Subsequently this product needs to be integrated over both $\delta\Omega_0$ and $\Omega_0$. We will have to separately analyse the symmetric and non-symmetric cases, considering the different conditions that need to be observed when generating anti-clockwise or clockwise movement in the rotor, as shown in Fig.~\ref{fig:NSConditons}.

\subsection{Symmetric Case}

The AVD can be expressed as:

\begin{eqnarray}
\hspace{-1.5cm}f\left(\Omega\right) & = & \Theta\left(-\Omega\right)\int_{-\infty}^{\infty}\textrm{d}\Omega_{0}\int_{-\infty}^{\infty}\textrm{d}\delta\Omega_{0}\, G(\delta\Omega_{0})H_{-}\left(\Omega|\Omega_{0}+\delta\Omega_0\right)f\left(\Omega_{0}\right)\nonumber \\
\hspace{-1.5cm}&  & +\Theta\left(\Omega\right)\int_{-\infty}^{\infty}\textrm{d}\Omega_{0}\int_{-\infty}^{\infty}\textrm{d}\delta\Omega_{0}\, G(\delta\Omega_{0})H_{+}\left(\Omega|\Omega_{0}+\delta\Omega_0\right)f\left(\Omega_{0}\right)
\label{eq:AVDapp}
\end{eqnarray}

In the symmetric case both terms in the AVD equation are equivalent since $f(\Omega) = f(-\Omega)$, and so we develop the equation only for positive $\Omega$, which corresponds to the second term in Eq.~\ref{eq:AVDapp},

\begin{eqnarray}
\hspace{-1.5cm}f\left(\Omega\right) & = & \int_{-\infty}^{\infty}\textrm{d}\Omega_{0}\int_{-\infty}^{\infty}\textrm{d}\delta\Omega_{0}\, G(\delta\Omega_{0})H_{+}\left(\Omega|\Omega_{0}+\delta\Omega_0\right)f\left(\Omega_{0}\right)\nonumber \\
\hspace{-1.5cm}& = & \int_{-\infty}^{\infty}\textrm{d}\Omega_{0}\int_{-\infty}^{\infty}\textrm{d}\delta\Omega_{0}\, G(\delta\Omega_{0})\Theta\left(\delta\Omega_{0}+\Omega_{0}-\Omega\right)\left|\frac{1}{\tau_c\dot\Omega}\right|\,\textrm{ei}\left(-\frac{t_\Omega}{\tau_c}\right)f\left(\Omega_{0}\right)\nonumber \\
\hspace{-1.5cm}& = & \int_{-\infty}^{\infty}\textrm{d}\Omega_{0}\int_{\Omega-\Omega_{0}}^{\infty}\textrm{d}\delta\Omega_{0}\, G(\delta\Omega_{0})\left|\frac{1}{\tau_c\dot\Omega}\right|\,\textrm{ei}\left(-\frac{t_\Omega}{\tau_c}\right)f\left(\Omega_{0}\right)\label{eq:NKRSym1}
\end{eqnarray}

Then, we can define the Kernel function 

\begin{equation}
  K\left(\Omega,\Omega_{0}\right)=\int_{\Omega-\Omega_{0}}^{\infty}\textrm{d}\delta\Omega_{0}\, G(\delta\Omega_{0})\left|\frac{1}{\tau_c\dot\Omega}\right|\,\textrm{ei}\left(-\frac{t_\Omega}{\tau_c}\right)
\end{equation}

$\fl\hspace{3.8cc}$and rewrite the AVD function as,

\begin{equation}
f(\Omega)=\int_{-\infty}^{\infty}\textrm{d}\Omega_{0}\, K\left(\Omega,\Omega_{0}\right)f\left(\Omega_{0}\right)
\end{equation}

Now we separate the integral in two parts, and considering that in the symmetric case the AVD follows $f(\Omega)=f(-\Omega)$, the distribution can be written as, 

\begin{eqnarray}
f\left(\Omega\right) & = & \int_{-\infty}^{0}\textrm{d}\Omega_{0}\, K\left(\Omega,\Omega_{0}\right)f\left(\Omega_{0}\right)+\int_{0}^{\infty}\textrm{d}\Omega_{0}\, K\left(\Omega,\Omega_{0}\right)f\left(\Omega_{0}\right)\nonumber \\
& = & \int_{0}^{\infty}\textrm{d}\Omega_{0}\, K\left(\Omega,-\Omega_{0}\right)f\left(-\Omega_{0}\right)+\int_{0}^{\infty}\textrm{d}\Omega_{0}\, K\left(\Omega,\Omega_{0}\right)f\left(\Omega_{0}\right)\nonumber \\
& = & \int_{0}^{\infty}\textrm{d}\Omega_{0}\, K\left(\Omega,-\Omega_{0}\right)f\left(\Omega_{0}\right)+\int_{0}^{\infty}\textrm{d}\Omega_{0}\, K\left(\Omega,\Omega_{0}\right)f\left(\Omega_{0}\right)\nonumber \\
& = & \int_{0}^{\infty}\textrm{d}\Omega_{0}\left(K\left(\Omega,-\Omega_{0}\right)+K\left(\Omega,\Omega_{0}\right)\right)f\left(\Omega_{0}\right)\nonumber \\
& = & \int_{0}^{\infty}\textrm{d}\Omega_{0}\,\widetilde{K}\left(\Omega,\Omega_{0}\right)f\left(\Omega_{0}\right)
\end{eqnarray}

\noindent where $\widetilde{K}\left(\Omega,\Omega_{0}\right)=K\left(\Omega,-\Omega_{0}\right)+K\left(\Omega,\Omega_{0}\right)$. There is no explicit analytic solution for the AVD, so in order to solve the integral equation in a numerical way, we write:

\begin{equation}
f\left(\Omega_{i}\right)  =  \sum_j\widetilde{K}\left(\Omega_{i},\Omega_{j}\right)f\left(\Omega_{j}\right)\Delta\Omega
\end{equation}

\noindent for a non-negative array of equidistant ($\Delta\omega$) points starting at $\Omega_1=0$, which we can put in a vector form

\begin{equation}
\vec{f}=\boldsymbol{\widetilde{K}}\cdot\vec{f}\quad,\,\widetilde{\textbf{K}}=[\,\widetilde{K}^{i,j}\,]\cdot\Delta\Omega
\end{equation}

We thus have a matrix eigenvalue problem to solve to find the AVD function.

\subsection{Non-Symmetric case}

\begin{figure}[h!]
  \begin{center}
   \includegraphics[scale=0.32]{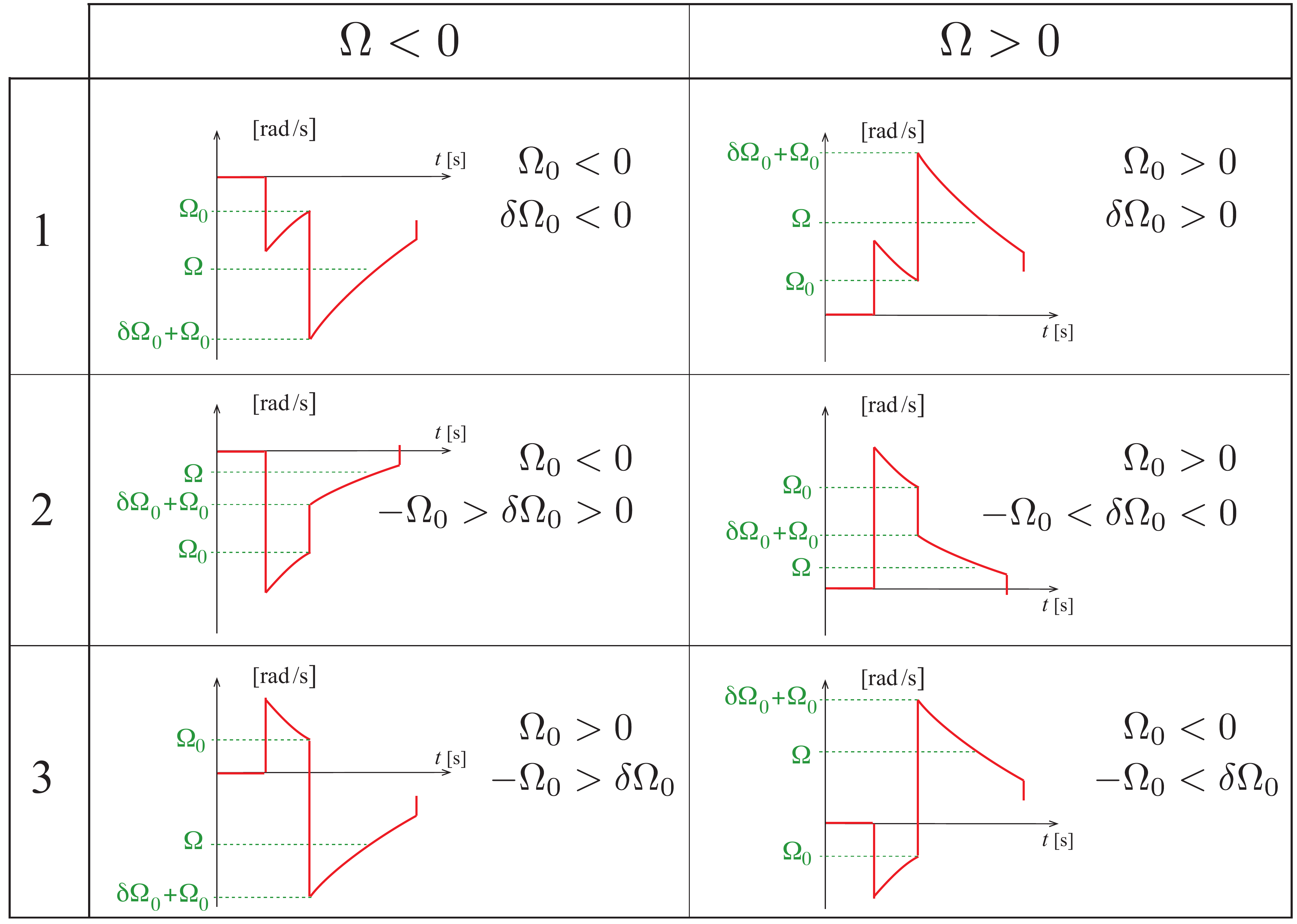}  
  \end{center}
  \caption{Schematic of the different conditions that will generate a clockwise or anti-clockwise movement in the rotor and that need to be treated separately in the analysis.}
  \label{fig:NSConditons}
  \end{figure}

When the rotor is asymmetric we have to consider additional conditions related to the direction of the kicks. Figure~\ref{fig:NSConditons} shows different ways to provoke rotor movement in the anti-clockwise direction (left) or in the clockwise one (right). Hence, for each direction the AVD has to be separated into three integrals, each of which corresponds to one of the three situations depicted in Fig.~\ref{fig:NSConditons}. Starting for the rotor moving in anti-clockwise direction ($\Omega<0$), we have

\begin{eqnarray}
\hspace{-2cm}f_{-}\left(\Omega\right) & = & \underbrace{\int_{-\infty}^{0}\mathrm{d}\Omega_{0}\int_{-\infty}^{0}\mathrm{d}\delta\Omega_{0}\, G_{-}\left(\delta\Omega_{0}\right)H_{-}\left(\Omega|\Omega_{0}+\delta\Omega_0\right)f_{-}\left(\Omega_{0}\right)}_{1}+\nonumber \\
\hspace{-2cm}&  & \underbrace{\int_{-\infty}^{0}\mathrm{d}\Omega_{0}\int_{0}^{-\Omega_{0}}\mathrm{d}\delta\Omega_{0}\, G_{+}\left(\delta\Omega_{0}\right)H_{-}\left(\Omega|\Omega_{0}+\delta\Omega_0\right)f_{-}\left(\Omega_{0}\right)}_{2}+\nonumber \\
\hspace{-2cm}&  & \underbrace{\int_{0}^{\infty}\mathrm{d}\Omega_{0}\int_{-\infty}^{-\Omega_{0}}\mathrm{d}\delta\Omega_{0}\, G_{-}\left(\delta\Omega_{0}\right)H_{-}\left(\Omega|\Omega_{0}+\delta\Omega_0\right)f_{+}\left(\Omega_{0}\right)}_{3}
\end{eqnarray}

\noindent and secondly for the rotor moving in clockwise direction ($\Omega>0$), we find

\begin{eqnarray}
\hspace{-2cm}f_{+}\left(\Omega\right) & = & \underbrace{\int_{0}^{\infty}\mathrm{d}\Omega_{0}\int_{0}^{\infty}\mathrm{d}\delta\Omega_{0}\, G_{+}\left(\delta\Omega_{0}\right)H_{+}\left(\Omega|\Omega_{0}+\delta\Omega_0\right)f_{+}\left(\Omega_{0}\right)}_{1}+\nonumber \\
\hspace{-2cm}&  & \underbrace{\int_{0}^{\infty}\mathrm{d}\Omega_{0}\int_{-\Omega_{0}}^{0}\mathrm{d}\delta\Omega_{0}\, G_{-}\left(\delta\Omega_{0}\right)H_{+}\left(\Omega|\Omega_{0}+\delta\Omega_0\right)f_{+}\left(\Omega_{0}\right)}_{2}+\nonumber \\
\hspace{-2cm}&  & \underbrace{\int_{-\infty}^{0}\mathrm{d}\Omega_{0}\int_{-\Omega_{0}}^{\infty}\mathrm{d}\delta\Omega_{0}\, G_{+}\left(\delta\Omega_{0}\right)H_{+}\left(\Omega|\Omega_{0}+\delta\Omega_0\right)f_{-}\left(\Omega_{0}\right)}_{3}
 \end{eqnarray}

\noindent (Note that $-\Omega_0>0$ in the last line.) Here the symbol $\pm$ in each function indicates if the angular velocity that it depends on is positive or negative respectively. Then, we can collect similar terms 

\begin{eqnarray}
\hspace{-4cm}f_{-}\left(\Omega\right) & = & \int_{-\infty}^{0}\mathrm{d}\Omega_{0}\left[\int_{-\infty}^{0}\mathrm{d}\delta\Omega_{0}\, G_{-}\left(\delta\Omega_{0}\right)H_{-}\left(\Omega|\Omega_{0}+\delta\Omega_0\right)  + \int_{0}^{-\Omega_{0}}\mathrm{d}\delta\Omega_{0}\, G_{+}\left(\delta\Omega_{0}\right)H_{-}\left(\Omega|\Omega_{0}+\delta\Omega_0\right)\right]f_{-}\left(\Omega_{0}\right)\nonumber \\
\hspace{-4cm}&  & +\int_{0}^{\infty}\mathrm{d}\Omega_{0}\int_{-\infty}^{-\Omega_{0}}\mathrm{d}\delta\Omega_{0}\, G_{-}\left(\delta\Omega_{0}\right)H_{-}\left(\Omega|\Omega_{0}+\delta\Omega_0\right)f_{+}\left(\Omega_{0}\right)\nonumber \\
\hspace{-4cm}& \equiv & \int_{-\infty}^{0}\mathrm{d}\Omega_{0}\, K_{1}\left(\Omega,\Omega_{0}\right)f_{-}\left(\Omega_{0}\right)+\int_{0}^{\infty}\mathrm{d}\Omega_{0}\, K_{2}\left(\Omega,\Omega_{0}\right)f_{+}\left(\Omega_{0}\right)
\end{eqnarray}

 \begin{eqnarray}
\hspace{-4cm}f_{+}\left(\Omega\right) & = & \int_{0}^{\infty}\mathrm{d}\Omega_{0}\left[\int_{0}^{\infty}\mathrm{d}\delta\Omega_{0}\, G_{+}\left(\delta\Omega_{0}\right)H_{+}\left(\Omega|\Omega_{0}+\delta\Omega_0\right)+\int_{-\Omega_{0}}^{0}\mathrm{d}\delta\Omega_{0}\, G_{-}\left(\delta\Omega_{0}\right)H_{+}\left(\Omega|\Omega_{0}+\delta\Omega_0\right)\right]f_{+}\left(\Omega_{0}\right)\nonumber \\
\hspace{-4cm}&  & +\int_{-\infty}^{0}\mathrm{d}\Omega_{0}\int_{-\Omega_{0}}^{\infty}\mathrm{d}\delta\Omega_{0}\, G_{+}\left(\delta\Omega_{0}\right)H_{+}\left(\Omega|\Omega_{0}+\delta\Omega_0\right)f_{-}\left(\Omega_{0}\right)\nonumber \\
\hspace{-4cm}& = & \int_{-\infty}^{0}\mathrm{d}\Omega_{0}\, K_{3}\left(\Omega,\Omega_{0}\right)f_{-}\left(\Omega_{0}\right)+\int_{0}^{\infty}\mathrm{d}\Omega_{0}\, K_{4}\left(\Omega,\Omega_{0}\right)f_{+}\left(\Omega_{0}\right)
\end{eqnarray}

To solve this set of integral equations numerically we again discretise de integrals 

\begin{eqnarray}
 \hspace{-2cm}f_-(\widetilde{\Omega}_i) &=& \sum_{j=-N}^{-1} 
  K_1\left(\widetilde{\Omega}_{i},\widetilde{\Omega}_{j}\right)\Delta\Omega\,\,f_-\left(\widetilde{\Omega}_{j}\right) +   
  \sum_{j=0}^N K_2\left(\widetilde{\Omega}_{i},\Omega_{j}\right)\Delta\Omega\,\,f_+\left(\Omega_{j}\right)\\
  \hspace{-2cm}f_+(\Omega_i) &=& \sum_{j=-N}^{-1}
  K_3\left(\Omega_{i},\widetilde{\Omega}_{j}\right)\Delta\Omega\,\,f_-\left(\widetilde{\Omega}_{j}\right) + 
  \sum_{j=0}^N K_4\left(\Omega_{i},\Omega_{j}\right)\Delta\Omega\,\,f_+\left(\Omega_{j}\right)  
\end{eqnarray}

\noindent where $[\widetilde{\Omega}_{-N},...,\widetilde{\Omega}_{-1}]$ is an array of $N$ negative values and $[\Omega_0,...,\Omega_N]$ an array of $N+1$ non-negative values (with $\Omega_0=0$) such that $\vec{f} \equiv [\widetilde{\Omega}_{-N},...,\widetilde{\Omega}_{-1},\Omega_0,...,\Omega_N]$ is an ordered array ($\Delta\Omega$) of equidistant values. Similar to the symmetric case, we can resolve the integral equations system as an eigenvalue problem,

\begin{equation}
\vec{f}=\mathrm{\mathbf{K}}\cdot\vec{f}\qquad,\:\mathrm{\mathbf{K}}=\left[\begin{array}{cc}
K_{1}^{i,j} & K_{2}^{i,j}\\
K_{3}^{i,j} & K_{4}^{i,j}
\end{array}\right]\cdot\Delta\Omega
\end{equation}

\noindent {where the solution, for the symmetric and asymmetric cases, corresponds to the eigenvector with eigenvalue 1.}
%%%%%%%%%%%%%%%%%%%%%%%%%%%%%%%%%%%%%%%%%%%%%%%%%%%%%%%%%%%%%%%%%%%%%%%%%%%%%%%%%%%%%%%%%%%%%%%%%%%%%%%%%%%%%%%%%%%%%%

%%%%%%%%%%%%%%%%%%%%%%%%%%%%%%%%%%%%%%%%%%%%%%%%%%%%%%%%%%%%%%%%%%%%%%%%%%%%%%%%%%%%%%%%%%%%%%%%%%%%%%%%%%%%%%%%%%%%%%
\section*{References}
\bibliographystyle{ieeetr}
\bibliography{GranularMotor}{}

\begin{thebibliography}{10}

\bibitem{smoluchowski_experimentell_1927}
M.~Smoluchowski, ``Experimentell nachweisbare, der üblichen {Thermodynamik}
  widersprechende {Molekularphänomene},'' {\em Pisma Mariana Smoluchowskiego},
  vol.~2, no.~1, pp.~226--251, 1927.

\bibitem{feynman_feynman_1977}
R.~P. Feynman, R.~B. Leighton, and M.~Sands, {\em The {Feynman} {Lectures} on
  {Physics}, {Vol}. 1: {Mainly} {Mechanics}, {Radiation}, and {Heat}}.
\newblock Reading, Mass.: Addison Wesley, 1 edition~ed., Feb. 1977.

\bibitem{reimann_brownian_2002}
P.~Reimann, ``Brownian motors: noisy transport far from equilibrium,'' {\em
  Physics Reports}, vol.~361, pp.~57--265, Apr. 2002.

\bibitem{talbot_effect_2012}
J.~Talbot and P.~Viot, ``Effect of dynamic and static friction on an asymmetric
  granular piston,'' {\em Physical Review E}, vol.~85, p.~021310, Feb. 2012.

\bibitem{sarracino_ratchet_2013}
A.~Sarracino, A.~Gnoli, and A.~Puglisi, ``Ratchet effect driven by {Coulomb}
  friction: the asymmetric {Rayleigh} piston,'' {\em arXiv:1303.0700
  [cond-mat]}, Mar. 2013.
\newblock arXiv: 1303.0700.

\bibitem{hanggi_brownian_2005}
P.~Hänggi, F.~Marchesoni, and F.~Nori, ``Brownian motors,'' {\em Annalen der
  Physik}, vol.~14, pp.~51--70, Feb. 2005.

\bibitem{eshuis_experimental_2010}
P.~Eshuis, K.~van~der Weele, D.~Lohse, and D.~van~der Meer, ``Experimental
  {Realization} of a {Rotational} {Ratchet} in a {Granular} {Gas},'' {\em
  Physical Review Letters}, vol.~104, p.~248001, June 2010.

\bibitem{balzan_brownian_2011}
R.~Balzan, F.~Dalton, V.~Loreto, A.~Petri, and G.~Pontuale, ``Brownian motor in
  a granular medium,'' {\em Physical Review E}, vol.~83, p.~031310, Mar. 2011.

\bibitem{talbot_kinetics_2011}
J.~Talbot, R.~D. Wildman, and P.~Viot, ``Kinetics of a {Frictional} {Granular}
  {Motor},'' {\em Physical Review Letters}, vol.~107, p.~138001, Sept. 2011.

\bibitem{gnoli_granular_2013}
A.~Gnoli, A.~Puglisi, and H.~Touchette, ``Granular {Brownian} motion with dry
  friction,'' {\em EPL (Europhysics Letters)}, vol.~102, p.~14002, Apr. 2013.

\bibitem{cleuren_dynamical_2008}
B.~Cleuren and R.~Eichhorn, ``Dynamical properties of granular rotors,'' {\em
  Journal of Statistical Mechanics: Theory and Experiment}, vol.~2008,
  p.~P10011, Oct. 2008.

\bibitem{talbot_kinetic_2011}
J.~Talbot, A.~Burdeau, and P.~Viot, ``Kinetic analysis of a chiral granular
  motor,'' {\em Journal of Statistical Mechanics: Theory and Experiment},
  vol.~2011, p.~P03009, Mar. 2011.

\bibitem{joubaud_fluctuation_2012}
S.~Joubaud, D.~Lohse, and D.~van~der Meer, ``Fluctuation {Theorems} for an
  {Asymmetric} {Rotor} in a {Granular} {Gas},'' {\em Physical Review Letters},
  vol.~108, p.~210604, May 2012.

\bibitem{kanazawa_asymptotic_2014}
K.~Kanazawa, T.~G. Sano, T.~Sagawa, and H.~Hayakawa, ``Asymptotic derivation of
  {Langevin}-like equation with non-{Gaussian} noise and its analytical
  solution,'' {\em arXiv:1412.2233 [cond-mat]}, Dec. 2014.
\newblock arXiv: 1412.2233.

\bibitem{costantini_models_2009}
G.~Costantini, A.~Puglisi, and U.~M.~B. Marconi, ``Models of granular
  ratchets,'' {\em Journal of Statistical Mechanics: Theory and Experiment},
  vol.~2009, p.~P07004, July 2009.

\bibitem{brilliantov_kinetic_2004}
N.~V. Brilliantov and T.~Poschel, {\em Kinetic {Theory} of {Granular} {Gases}}.
\newblock Oxford ; New York: Oxford University Press, Sept. 2004.

\bibitem{atkinson_numerical_2009}
K.~E. Atkinson, {\em The {Numerical} {Solution} of {Integral} {Equations} of
  the {Second} {Kind}}.
\newblock Cambridge; New York: Cambridge University Press, reissue edition~ed.,
  Mar. 2009.

\end{thebibliography}

\end{document}